\documentclass[lettersize,journal]{IEEEtran}
\usepackage{amsmath,amsfonts}
\usepackage{algorithmic}
\usepackage{algorithm}
\usepackage{array}
\usepackage{textcomp}
\usepackage{stfloats}
\usepackage{url}
\usepackage{verbatim}
\usepackage{graphicx}
\usepackage{cite}
\usepackage{booktabs}
\usepackage{mathtools}
\usepackage{bbm}
\usepackage{bm}
\usepackage{bbding}
\usepackage{amssymb}
\usepackage{multirow}
\usepackage[colorlinks=true,linkcolor=blue,citecolor=green,urlcolor=blue,]{hyperref}
\usepackage[caption=false,font=footnotesize,labelfont=rm,textfont=rm]{subfig}

\let\sss= \scriptscriptstyle

\begin{document}

\title{Game-Theoretic Modeling of Vehicle Unprotected Left Turns Considering Drivers’ Bounded Rationality}

\author{Yuansheng Lian, Ke Zhang, Meng Li, Shen Li

\thanks{This work was supported by grants from National Key Research and Development Program of China (2022YFB2503200), Tsinghua University-Mercedes Benz Joint Institute for Sustainable Mobility. \textit{(Corresponding author: Meng Li.)}

Meng Li is with the Department of Civil Engineering, Tsinghua University, Beijing 100084, China, and also with the State Key Laboratory of Intelligent Green Vehicle and Mobility, Tsinghua University, Beijing 100084, China (e-mail: mengli@tsinghua.edu.cn)

Yuansheng Lian, Ke Zhang, and Shen Li are with the Department of Civil Engineering, Tsinghua University, Beijing 100084, China (e-mail: lys22@mails.tsinghua.edu.cn; zhangkethu@mail.tsinghua.edu.cn; sli299@tsinghua.edu.cn).}}


\maketitle

\begin{abstract}
Modeling the decision-making behavior of vehicles presents unique challenges, particularly during unprotected left turns at intersections, where the uncertainty of human drivers is especially pronounced.
In this context, connected autonomous vehicle (CAV) technology emerges as a promising avenue for effectively managing such interactions while ensuring safety and efficiency.
Traditional approaches, often grounded in game theory assumptions of perfect rationality, may inadequately capture the complexities of real-world scenarios and drivers' decision-making errors.
To fill this gap, we propose a novel decision-making model for vehicle unprotected left-turn scenarios, integrating game theory with considerations for drivers' bounded rationality. Our model, formulated as a two-player normal-form game solved by a quantal response equilibrium (QRE), offers a more nuanced depiction of driver decision-making processes compared to Nash equilibrium (NE) models. Leveraging an Expectation-Maximization (EM) algorithm coupled with a subtle neural network trained on precise microscopic vehicle trajectory data, we optimize model parameters to accurately reflect drivers' interaction-aware bounded rationality and driving styles.
Through comprehensive simulation experiments, we demonstrate the efficacy of our proposed model in capturing the interaction-aware bounded rationality and decision tendencies between players. 
The proposed model proves to be more realistic and efficient than NE models in unprotected left-turn scenarios. 
Our findings contribute valuable insights into the vehicle decision-making behaviors with bounded rationality, thereby informing the development of more robust and realistic autonomous driving systems.
\end{abstract}


\section{Introduction}
Connected autonomous vehicle (CAV) refers to a vehicle that can operate autonomously and communicate with other vehicles and infrastructure to enhance safety and efficiency.
It is widely acknowledged that, in the foreseeable future, CAVs will co-exist with human-driven vehicles (HDVs) in traffic \cite{li2020game}, necessitating continuous interactions between the two. Consequently, there arises an urgent need to develop models that enable the operation of CAVs within mixed traffic environments, enabling them to anticipate the intentions of surrounding human drivers and make human-like decisions based on these expectations and feedback.

In the context of mixed traffic environments, one of the most prevalent scenarios entails vehicles executing unprotected left turns at signalized intersections. 
According to the data from New York City Department of Transportation \cite{choi2010crash}, left-turns is a contributing factor in approximately 61\% of all accidents that happened when a vehicle is crossing an intersection during 2005 and 2007. In addition, left-turn accidents often result in severe injuries or fatalities.
However, modeling the decision-making behavior of vehicles taking unprotected left turns presents a formidable challenge, attributing to the intricate complexities in human driving behaviors.

The sequencing of left-turn maneuvers at intersections is normally based on the availability of gaps between through vehicles and the willingness of left-turning vehicles to exploit such openings. However, real-world observations have shown that the driver behavior in this scenario can be more complex. For instance, through vehicles may adjust their speed to impede left-turning vehicles or display courtesy by decelerating to accommodate them \cite{wang2019enabling}. In addition, human drivers at intersections can exhibit unsafe and erratic turning behaviors or experience significant reductions in driving speeds due to factors such as lack of awareness, delayed reactions, and mis-decision-making. The failure to judge the existence of a sufficient gap in the opposing traffic, misjudgment of the opposing vehicle's speed and intention, and inadequate or obstructed sight distance may lead to unpredictable crashes.

CAVs offer several advantages over human drivers, including precise monitoring of distance and instantaneous responses to diverse driving situations. By understanding driver behaviors within such scenarios, CAVs can enhance their decision-making processes, thus executing safer trajectories at intersections. Notably, researchers have increasingly turned to game theory—a valuable tool for mathematically modeling strategic interactions among rational agents—to model the interactive behavior of CAVs across various traffic scenarios \cite{li2020game, ali2019game, lopez2022game, wang2021competitive, rahmati2017towards, rahmati2021helping}.

In game-theoretic modeling, the optimal decisions of players are often derived through  Nash Equilibrium (NE), where each player selects strategies to maximize expected payoffs \cite{qin2024game}. It is based on two assumptions: 1) every player has perfect rationality that allows them to take actions that maximizes their payoff; and 2) each player completely understands other players' decision-making processes, so they can predict the course of actions other drivers will take. The existing literature \cite{rahmati2017towards, rahmati2021helping} on vehicle left-turn decision-making scenarios mainly adopts this setting.
However, in the context of unprotected left-turn scenarios, drivers are not perfect optimizers in the complex driving environment due to the limitations of computational and cognitive skills. The rationality of drivers in left-turns is bounded, leading to unforeseen deviations from expected outcomes and highlighting the need to account for drivers' bounded rationality in decision-making processes \cite{sivak2002common}, a consideration largely absent in existing literature.

To address this gap, this study models vehicle unprotected left-turn behavior as a two-player game considering drivers' bounded rationality. By formulating the behavior as a two-player game and deriving a logit-form quantal response equilibrium, we seek to capture the nuanced decision-making processes underlying such scenarios. High-precision vehicle trajectory data are used to calibrate the model and optimize the parameters. Simulation experiments are conducted to evaluate the performance of the proposed model.

The main contribution of this study is threefold:
\begin{itemize}
    \item We develop a game-theoretic decision-making framework to model unprotected left-turn maneuvers of vehicles, which elaborately incorporates an interaction-aware bounded rationality and drivers' decision-making tendencies based on a quantal response equilibrium (QRE).
    \item We design an Expectation–Maximization (EM) algorithm to calibrate the game model and solve the equilibrium based on real-world vehicle trajectory data.
    A subtle neural network is introduced to simultaneously learn the payoff weights with decision tendencies and interaction-aware bounded rationality values during the iteration of EM algorithm.
    \item We conduct simulation experiments to validate the proposed model. The performance of the proposed model is compared with two baseline models to assess its superiority. 
\end{itemize}

The remainder of this paper is organized as follows. 
Section \ref{sec:lr} reviews the related literature. Section \ref{sec:pf} formulates the game model. 
And then, Section \ref{sec:brm} introduces the drivers' bounded rationality modeling approaches. Simulation experimental settings are presented in Section \ref{sec:es} and followed by the experimental results in Section \ref{sec:r}. Section \ref{sec:c} concludes the study and indicates some future research directions.

\section{Literature Review}\label{sec:lr}





\subsection{Vehicles Decision-Making Modeling at Intersections}
Numerous models have been proposed to handle vehicle interaction at intersections in a CAV environment, where vehicles interact with each other and cooperate with road infrastructure to resolve traffic conflicts.
Vehicles may cooperate through vehicle-to-vehicle (V2V) negotiations \cite{ahmane2013modeling, de2013autonomous} or be controlled by a centralized traffic manager \cite{lee2012development, chen2021rhythmic, lin2021rhythmic, liu2024integrated}. 
Although strategies based on cooperative driving have been shown to be capable of improving intersection traffic safety and efficiency, they rely on strong assumptions that penetration rate of CAV is considerably high, and vehicle-to-vehicle or vehicle-to-infrastructure communications are well developed, which will not be the case in the near term.

Alternative strategies have been focused on individualized control of the CAV. To account for the interactions among vehicles, strategies based on optimization \cite{schwarting2017safe, basil2023evaluation},  
learning \cite{you2019advanced, liu2023towards}, and game theory \cite{li2020game, wang2021competitive, rahmati2021helping, liu2022three, fang2024cooperative}, have been proposed.
Optimization-based models aim to generate desirable trajectories with frameworks such as model predictive control and Bezier curve optimization, but neglecting the dynamic interaction of vehicles in the complex traffic environment \cite{niels2024optimization}. 
More recently, learning-based methods have gained much attention owing to advancements in artificial intelligence (AI), with vehicles decision-making approaches with supervised learning \cite{li2018humanlike, wang2024improving} and reinforcement learning \cite{liu2023towards, hou2024merging} emerging prominently.
However, these learning-based methods face challenges including difficulty in acquiring training data, heavy training burden, and poor interpretability \cite{lu2023game}.

Game theory has proven to be a valuable tool for modeling strategic interactions among intelligent vehicles and can be classified into cooperative games and non-cooperative games. Cooperative games assume that vehicles work towards a shared goal through V2V communication \cite{jing2019cooperative}. Conversely, in non-cooperative games, vehicles are dedicated to optimize their individual gains.
In literature \cite{mandiau2008behaviour}, the vehicle interactions at an intersection were modeled using normal-form games. Vehicles select actions between “Stop” and “Go” according to their payoff functions. However, the performance of the approach was limited due to the limited number of action choices (i.e., two).
In literature \cite{sadigh2016planning}, an improvement was made taking into account the dynamics of the vehicle. The experimental results of an intersection scenario with two interacting vehicles both driving straight to cross the intersection were reported. 
Li et al. \cite{li2020game} model the interactive decision-making processes of vehicles at uncontrolled intersections based on an extensive-game formulation. Additional modeling considerations include courteous driving, limited perception range, and the breaking of deadlocks via exploratory actions. 
Different driving styles and social interaction characteristics are formulated for CAVs with respect to driving safety, ride comfort, and travel efficiency \cite{hang2020human, lu2023game, jia2023interactive}. 
Some studies further extend the game formulation to multi-vehicle interaction scenarios \cite{yan2023multi}.

The determination of payoff weights in the game is critical and requires manual design and experiments. To this end, some data-driven methods have been proposed to learn the game parameters.
In literature \cite{liu2022three}, a three-level game-theoretical modeling framework is proposed to generate safe and effective decisions for autonomous vehicles. The whole simulation procedure is divided into offline and online operations. At the offline stage, a neural network is used to learn the parameters via supervised learning.  At the online stage, real-time decisions are made based on re-evaluation and data acquisition periodically. 
Some studies incorporate reinforcement learning (RL) to learn the payoff weights \cite{yuan2021deep, liu2023towards}.

\subsection{Vehicle Unprotected Left-Turn Modeling}

In urban signalized intersections, the interaction between unprotected left-turning vehicles and opposing through vehicles represents a common scenario of traffic conflict streams \cite{shen2023analysis}. Numerous studies have explored the decision-making modeling for unprotected left-turning vehicles in a connected environment.

In study \cite{rahmati2017towards}, Rahmati et al. consider the problem of unprotected left turn in mixed environment. The interactive behavior among involved drivers was characterized using the theory of Nash equilibrium in non-cooperative simultaneous move games.
A following study \cite{rahmati2021helping} expands on the idea and proposes a different decision-making model using the sequential move structure in Stackelberg games. Each player in the game is presented with two strategies to choose from, i.e., turn \& wait (left-turning vehicle), keep driving \& decelerate (through vehicle). The decision modeling framework accounts for the behavioral uncertainties by defining the payoff function to be a combination of deterministic term and an error term. 

Researchers also integrate the task of decision making and motion planning of left-turning vehicles.
Zhou et al. \cite{zhou2019autonomous} propose a decision-plan-action framework combined with the estimation of interacting agents to plan the left-turn motion at mixed-flow intersections. The decision making and planning modules are realized through a binary logit model and Bezier curve optimization. A following study \cite{zhou2022autonomous} further considers the intended cooperation in left turn scenarios. Related studies also consider the path variations in the unprotected left turns \cite{zhao2023unprotected, liu2023teaching}. In \cite{liu2023teaching}, a left-turn planning framework is proposed to replicate human driving patterns with social intent. Maximum entropy inverse reinforcement learning (ME-IRL) is utlized to assess human trajectory preferences, and a Boltzmann distribution-based method is proposed to select optimal trajectory.

\subsection{Bounded Rationality Modeling in Vehicle Decision-Making}
Standard Nash equilibrium assumes that the players are rational decision-makers capable of assessing their set of available options and selecting the strategy that yields the most preferred outcome. Due to these assumptions, the Nash equilibrium concept often yields sharp theoretical predictions \cite{wang2022modeling}. Moreover, in the real-world decision-making process of vehicles, the phenomenon of perfect rationality rarely exists, as complexities and various factors can influence their choices.

Bounded Rationality refers to a cognitive limitation where human decision-making aims to satisfy rather than optimize, acknowledging the constraints and limitations in processing information and making choices.
McKelvey and Palfrey \cite{mckelvey1995quantal} first defined a quantal response equilibrium (QRE) of a normal-from game considering players' bounded rationality. A logit specification of the error structure was also derived.
Although the QRE framework has already been thoroughly investigated in behavioral economics, it is relatively new in the literature on vehicle decision making.

Wang et al. \cite{wang2022modeling} developed a QRE-based game-theoretic model in discretionary lane change scenarios. The bounded rationality in the model exponentially decays over decision time. Data from the Next Generation Simulation (NGSIM) program \cite{alexiadis2004next} were used to calibrate the model parameters. The model prediction accuracy reached 91.5\% and outperformed Talebpour’s game model \cite{talebpour2015modeling}.
Study \cite{chen2023game} also considers the bounded rationality in lane change interactions on highway on-ramps.


In summary, pioneering studies have explored the application of Quantal Response Equilibrium (QRE) in the decision-making processes of traffic agents. However, the majority of these studies assume that bounded rationality is either fixed \cite{chen2023game} or decays over time according to a specific function \cite{wang2022modeling}. In our study, we propose to model the interaction-aware bounded rationality in left-turn scenarios as trainable parameters in neural networks. Furthermore, existing research predominantly models left-turn decision-making at intersections at a coarse-grained level, focusing on whether to make a left turn or not. A more fine-grained model is needed that considers vehicle dynamics throughout the entire left-turn decision-making process.

\section{Problem Formulation} \label{sec:pf}
In our study, a two-player nonzero-sum normal-form game is established to model vehicle unprotected left-turn behavior at intersections.
The underlying reasons for just considering the game between two vehicles lies in the interaction mechanism in vehicle left-turn scenario.
When a left-turning vehicle decides to yield (or go) at the unprotected left-turn phase, the first gap in the opposing traffic and the intention of opposing leading vehicle are typically taken into consideration.
Hence,  a sub-game between the left-turning vehicle and the nearest approaching vehicle is triggered when we consider the whole opposing traffic.
The behavior of the following opposing vehicles can be represented with car-following models. 
As a result, the game primarily involves two vehicles nearest to the conflict point. The multi-vehicle interaction can be split into multiple two-player games, using pairwise comparison.

\subsection{Player and strategy}

\begin{figure}[t]
  \centering  
  {\includegraphics[width=\linewidth]{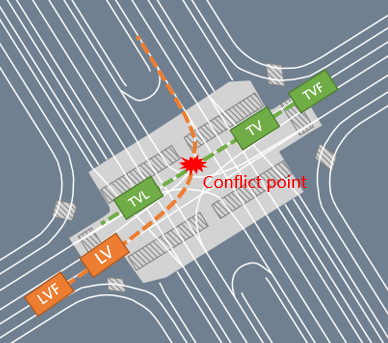}
  \caption{Interaction between left-turning vehicle and through vehicles. TV, TVL and TVF indicates through vehicle, leading through vehicle and following through vehicle, respectively. LV and LVF indicates left-turning vehicle and following left-turning vehicle, respectively. In this study, we only consider the interaction of TV and LV.}
  \label{fig:int-lf-th}}
\end{figure}

Figure \ref{fig:int-lf-th} displays a typical unprotected left-turn scenario.
In this study, two players are considered, including a left-turning vehicle (LV) and a through vehicle (TV).
Both players are required to be the leading vehicle to the potential conflict point.

In the two-player game, let $\gamma^{\mathrm{\sss LV}}$ and $\gamma^{\mathrm{\sss TV}}$ denote the actions of the left-turning vehicle and through vehicle, which are the longitudinal accelerations of the vehicles. Each player takes its action value from the finite action space $\Gamma^{\mathrm{\sss LV}}$ and  $\Gamma^{\mathrm{\sss TV}}$ accordingly as presented below.  
\begin{align}
    &\Gamma^{\mathrm{\sss LV}} = \{-1, 0, 1\}, m/s^2 \\
    &\Gamma^{\mathrm{\sss TV}} = \{-2, -1, 0, 1, 2\}, m/s^2 
\end{align}
In line with the prevailing approach in existing literature \cite{li2020game, ali2019game, rahmati2021helping, chen2023game}, a discrete action set instead of a continuous one is adopted in this study.
In the decision-making module, we only consider high-level longitudinal decisions in the current game model. 

\subsection{Payoff Function}
A vehicle $i$ makes action to maximize or attempt to optimize its own payoff function given the current states and actions, denoted by $J^{i}(\gamma^{\mathrm{\sss LV}}, \gamma^{\mathrm{\sss TV}})$, $i\in\{\mathrm{LV}, \mathrm{TV}\}$.
Three performance indexes are taken into consideration in decision-making modeling, i.e., safety, efficiency, and traffic rules. 
The payoff function is the weighted sum of these three components:
\begin{align}\label{eq:payoff}
    J^i(\gamma^{\mathrm{\sss LV}}, \gamma^{\mathrm{\sss TV}}) = 
    \omega_{s}^i(\gamma^{\mathrm{\sss LV}}, \gamma^{\mathrm{\sss TV}}) \cdot J_s^i(\gamma^{\mathrm{\sss LV}}, \gamma^{\mathrm{\sss TV}})&& \nonumber \\+ \
    \omega_e^i(\gamma^{\mathrm{\sss LV}}, \gamma^{\mathrm{\sss TV}}) \cdot  J_e^i(\gamma^{\mathrm{\sss LV}}, \gamma^{\mathrm{\sss TV}}) && \nonumber \\
    + \
    \omega_r^i(\gamma^{\mathrm{\sss LV}}, \gamma^{\mathrm{\sss TV}}) \cdot  J_r^i(\gamma^{\mathrm{\sss LV}}, \gamma^{\mathrm{\sss TV}}) && \nonumber \\
    i\in \{\mathrm{LV}, \mathrm{TV}\} &&
\end{align}

In Equation \ref{eq:payoff}, $\omega_s^i(\gamma^{\mathrm{\sss LV}}, \gamma^{\mathrm{\sss TV}})$, $\omega_e^i(\gamma^{\mathrm{\sss LV}}, \gamma^{\mathrm{\sss TV}})$ and $\omega_r^i(\gamma^{\mathrm{\sss LV}}, \gamma^{\mathrm{\sss TV}})$ are the corresponding weights for payoff $J_s^i$, $J_s^i$ and $J_r^i$ when vehicle $i$ takes action $\gamma^{\mathrm{\sss LV}}$ and vehicle TV takes action $\gamma^{\mathrm{\sss TV}}$. 
Most studies assume fixed payoff weights across different actions of both players, except for variations among three performance indexes. However, we argue that drivers have divergent senses of safety, efficiency, and traffic rules when taking different actions (i.e., accelerating, decelerating, or keeping speed). 
Hence, in this study, payoff weights considering drivers' decision tendency are adopted by allowing weights to varies across actions of players. However, the following constraints should always be satisfied.
\begin{align*}
    &\omega_s^i(\gamma^{\mathrm{\sss LV}}, \gamma^{\mathrm{\sss TV}}) + \omega_e^i(\gamma^{\mathrm{\sss LV}}, \gamma^{\mathrm{\sss TV}}) + \omega_r^i(\gamma^{\mathrm{\sss LV}}, \gamma^{\mathrm{\sss TV}}) = 1 \\
&\omega_s^i(\gamma^{\mathrm{\sss LV}}, \gamma^{\mathrm{\sss TV}}) \ge 0 \\
&\omega_e^i(\gamma^{\mathrm{\sss LV}}, \gamma^{\mathrm{\sss TV}}) \ge 0 \\
&\omega_r^i(\gamma^{\mathrm{\sss LV}}, \gamma^{\mathrm{\sss TV}}) \ge 0 
\end{align*}
where $i\in \{\mathrm{LV}, \mathrm{TV}\}, \gamma^{\mathrm{\sss LV}} \in \Gamma^{\mathrm{\sss LV}}, \gamma^{\mathrm{\sss TV}} \in \Gamma^{\mathrm{\sss TV}}$. 
To simplify, payoff weights in the following content will be referred to as $\omega_s^i \in \mathbb{R}^{|\Gamma^{\mathrm{\sss LV}}| \times |\Gamma^{\mathrm{\sss TV}}|}$, $\omega_e^i \in \mathbb{R}^{|\Gamma^{\mathrm{\sss LV}}| \times |\Gamma^{\mathrm{\sss TV}}|}$, $\omega_r^i \in \mathbb{R}^{|\Gamma^{\mathrm{\sss LV}}| \times |\Gamma^{\mathrm{\sss TV}}|}$ or $\omega^i = [\omega_s^i, \omega_e^i, \omega_r^i] \in \mathbb{R}^{3 \times |\Gamma^{\mathrm{\sss LV}}| \times |\Gamma^{\mathrm{\sss TV}}|}$, $i\in \{\mathrm{LV}, \mathrm{TV}\}$, in a matrix form or a tensor form. 

\subsubsection{Safety based payoff}
It is reasoned that the perceived closeness of oncoming traffic relative to the point of conflict, instead of the stop bar, is more important for drivers’ gap acceptance \cite{chan2006characterization}.

We use the sum of time to the collision point (TTCP) and relative time to collision (RTTC) \cite{chen2017surrogate} to measure the real-time safety risk.
TTCP is defined as the time to the potential collision point if the ego vehicle would not change its speed and follow the current path.
In addition, RTTC represents the time difference between the first vehicle that reaches the potential conflict point and the second driver arriving at the same location if they maintain their current speeds and paths. Lower TTCP and RTTC indicate a more urgent case with a higher possibility of collision.
\begin{align}
    &J_{s}^{\mathrm{\sss LV}} =  TTCP^{\mathrm{\sss LV}} + RTTC^{\mathrm{\sss LV}} \\
    &J_{s}^{\mathrm{\sss TV}} =  TTCP^{\mathrm{\sss TV}} + RTTC^{\mathrm{\sss TV}} \\
    &TTCP^{\mathrm{\sss LV}} = \frac{d_{\mathrm{conf}}^{\mathrm{\sss LV}}}{v^{\mathrm{\sss LV}}} \\
    &TTCP^{\mathrm{\sss TV}} = \frac{d_{\mathrm{conf}}^{\mathrm{\sss TV}}}{v^{\mathrm{\sss TV}}} \\
     &RTTC^{\mathrm{\sss LV}} = |TTCP^{\mathrm{\sss LV}} - TTCP^{\mathrm{\sss TV}}|\\ 
     &RTTC^{\mathrm{\sss TV}} = |TTCP^{\mathrm{\sss TV}} - TTCP^{\mathrm{\sss LV}}|  
\end{align}
where $d_{\mathrm{conf}}^{i}$ is the distance from the head of vehicle $i$ to the conflict point following the pre-planned path,
$v^i$ is the longitudinal velocity of vehicle $i$, $i \in \{\mathrm{LV}, \mathrm{TV}\}$.

\subsubsection{Efficiency based payoff}

We use the negative value of the time to reach destination (target exit lane) along the pre-planned path as the efficiency based payoff. Higher efficiency or shorter time to reach destination is encouraged.
\begin{align}
    &J_{e}^{\mathrm{\sss LV}} = -\frac{L_{des}^{\mathrm{\sss LV}}}{v^{\mathrm{\sss LV}}}\\
    &J_{e}^{\mathrm{\sss TV}} = -\frac{L_{des}^{\mathrm{\sss TV}}}{v^{\mathrm{\sss TV}}}
\end{align}
where $L_{des}^{i}$ is the distance to the destination of vehicle $i$. 

\subsubsection{Rule based payoff}

In vehicles decision-making, by granting the leader some advantages over the follower, it is possible to represent certain superior aspects of one side over the other in real-world competitive situations \cite{li2020game}.
For the rule based payoff, we grant the through vehicle with the superiority through a higher constant value. 
\begin{align}
    &J_r^{\mathrm{\sss LV}} = 0.5\\
    &J_r^{\mathrm{\sss TV}} = 1
\end{align}

\subsection{Game model}
With the actions and payoff functions defined in the previous sub-sections, we construct the two-player game in a tabular form, as shown in Table \ref{tab:game}.
$J_{m,n}^{i}$ denotes $J^i(\gamma^{\mathrm{\sss LV}} = m, \gamma^{\mathrm{\sss TV}} = n)$ for vehicle $i$, $i \in \{\mathrm{LV}, \mathrm{TV}\}$.

\begin{table*}[thb]
\caption{The structure of the normal-form game between left-turning vehicle (LV) and through vehicle (TV)}\label{tab:game}
\centering
\begin{tabular}{ccccccc}
\toprule
                     &    & \multicolumn{5}{c}{$\gamma^{\mathrm{\sss TV}}$ (m/s$^2$)} \\ \cline{3-7} 
                     &   &-2  & -1        & 0    & 1  & 2  \\\midrule
\multirow{3}{*}{$\gamma^{\mathrm{\sss LV}}$(m/s$^2$)} & -1 & ($J_{-1,-2}^{\mathrm{\sss LV}}$, $J_{-1,-2}^{\mathrm{\sss TV}}$)    & ($J_{-1,-1}^{\mathrm{\sss LV}}$, $J_{-1,-1}^{\mathrm{\sss TV}}$)    &  ($J_{-1,0}^{\mathrm{\sss LV}}$, $J_{-1,0}^{\mathrm{\sss TV}}$)    &  ($J_{-1,1}^{\mathrm{\sss LV}}$, $J_{-1,1}^{\mathrm{\sss TV}}$) &  ($J_{-1,2}^{\mathrm{\sss LV}}$, $J_{-1,2}^{\mathrm{\sss TV}}$)   \\
                     & 0  &  ($J_{0,-2}^{\mathrm{\sss LV}}$, $J_{0,-2}^{\mathrm{\sss TV}}$)         &($J_{0,-1}^{\mathrm{\sss LV}}$, $J_{0,-1}^{\mathrm{\sss TV}}$)         & ($J_{0,0}^{\mathrm{\sss LV}}$, $J_{0,0}^{\mathrm{\sss TV}}$)     & ($J_{0,1}^{\mathrm{\sss LV}}$, $J_{0,1}^{\mathrm{\sss TV}}$)  & ($J_{0,2}^{\mathrm{\sss LV}}$, $J_{0,2}^{\mathrm{\sss TV}}$)     \\
                     & 1  & ($J_{1,-2}^{\mathrm{\sss LV}}$, $J_{1,-2}^{\mathrm{\sss TV}}$)          &($J_{1,-1}^{\mathrm{\sss LV}}$, $J_{1,-1}^{\mathrm{\sss TV}}$)          &   ($J_{1,0}^{\mathrm{\sss LV}}$, $J_{1,0}^{\mathrm{\sss TV}}$)   &($J_{1,1}^{\mathrm{\sss LV}}$, $J_{1,1}^{\mathrm{\sss TV}}$)&($J_{1,2}^{\mathrm{\sss LV}}$, $J_{1,2}^{\mathrm{\sss TV}}$) \\
\bottomrule
\end{tabular}
\end{table*}

\section{Bounded rationality modeling} \label{sec:brm}
The rationality theory is a fundamental component of game theory. It posits that decision-making rationality is bounded or limited for individuals.
Under these limitations, rational individuals will tend to choose a satisfactory decision instead of an optimal one.
McKelvey and Palfrey \cite{mckelvey1995quantal} were pioneers in investigating the application of standard statistical models to quantal choice in a game-theoretic setting and defining a quantal response equilibrium. 
In this study, we define a QRE in unprotected left-turn scenario and develop an equilibrium solution, as outlined in the following subsections.

\subsection{Quantal response equilibrium}
Let $J^{\mathrm{\sss LV}}$ and $J^{\mathrm{\sss LV}}$ denote the payoffs of the left-turning vehicle and the through vehicle.
In the QRE, drivers' action is a random variable defined over the action set.
For the left-turning vehicle, each allowable action $\gamma_i^{\mathrm{\sss LV}}$ is selected with probability $p_i^{\mathrm{\sss LV}}$. For the through vehicle, each allowable action $\gamma_i^{\mathrm{\sss TV}}$ is selected with probability $p_i^{\mathrm{\sss TV}}$. 
In this study, we adopt a logit-form choice model to calculate $p_i^{\mathrm{\sss LV}}$ and $p_i^{\mathrm{\sss TV}}$ as follows.

\begin{align}\label{eq:pq}
    p_i^{\mathrm{\sss LV}} &= \frac{exp\{\lambda^{\mathrm{\sss LV}} E[J^{\mathrm{\sss LV}}(\gamma_i^{\mathrm{\sss LV}})]\}}{\sum_{\gamma_j^{\mathrm{\sss LV}}  \in \Gamma^{\mathrm{\sss LV}}} exp\{\lambda^{\mathrm{\sss LV}} E[J^{\mathrm{\sss LV}}(\gamma_j^{\mathrm{\sss LV}})]\}} \\
    p_i^{\mathrm{\sss TV}} &= \frac{exp\{\lambda^{\mathrm{\sss TV}} E[J^{\mathrm{\sss TV}}(\gamma_i^{\mathrm{\sss TV}})]\}}{\sum_{\gamma_j^{\mathrm{\sss TV}}  \in \Gamma^{\mathrm{\sss TV}}} exp\{\lambda^{\mathrm{\sss TV}} E[J^{\mathrm{\sss TV}}(\gamma_j^{\mathrm{\sss TV}})]\}}\label{eq:pq2}
\end{align}
where $E[J^{\mathrm{\sss LV}}(\gamma_i^{\mathrm{\sss LV}})]$ denotes the expected payoffs of LV choosing $\gamma_i^{\mathrm{\sss LV}}$ and $E[J^{\mathrm{\sss TV}}(\gamma_i^{\mathrm{\sss TV}})]$ denotes the expected payoffs of TV choosing $\gamma_i^{\mathrm{\sss TV}}$.
\begin{align}\label{eq:e1}
    E[J^{\mathrm{\sss LV}}(\gamma_i^{\mathrm{\sss LV}})] &= \sum_{\gamma_j^{\mathrm{\sss TV}}  \in \Gamma^{\mathrm{\sss TV}}} p_j^{\mathrm{\sss TV}} \times J^{\mathrm{\sss LV}}(\gamma_i^{\mathrm{\sss LV}}, \gamma_j^{\mathrm{\sss LV}}) \\ 
    E[J^{\mathrm{\sss TV}}(\gamma_i^{\mathrm{\sss TV}})] &= \sum_{\gamma_j^{\mathrm{\sss LV}}  \in \Gamma^{\mathrm{\sss LV}}} p_j^{\mathrm{\sss LV}} \times J^{\mathrm{\sss TV}}(\gamma_j^{\mathrm{\sss LV}}, \gamma_i^{\mathrm{\sss LV}}) \label{eq:e2}
\end{align}
Here, 
$\lambda^i$ ($i \in \{\mathrm{LV}, \mathrm{TV}\}$)
in Equation \ref{eq:pq} and \ref{eq:pq2} is the bounded rationality value inversely related to the level of error: $\lambda^i = 0$ indicates that actions consist of all error, and $\lambda^i \to \inf$ indicates that there is no error and the QRE approaches Nash equilibrium \cite{mckelvey1995quantal}.

Related studies normally assume that the bounded rationality is fixed \cite{chen2023game} or changes exponentially over time \cite{wang2022modeling}. However, in vehicle unprotected left-turn scenarios, we argue that the drivers' rationality is primarily influenced by the drivers' sense of collision risk and interacting behavior.
In the proposed model, the bounded rationality value $\lambda^i$ is obtained from a learnable vector $\bm{\lambda}^i \in \mathbb{R}^{d_r}$. 
Each element of $\bm{\lambda}^i$ corresponds to the bounded rationality value of player $i$ at a specific discretized distance to the conflict point $d_{\mathrm{conf}}^{i}$. Given $\bm{\lambda}^i$, bounded rationality value $\lambda^i$ can be calculated by: 
\begin{align}
    \lambda^i = \sum_{j=1}^{d_r} \mathbf{1}(d = d_j) \cdot \bm{\lambda}^i_j
\end{align}
where \(\mathbf{1}(d = d_j)\) is the indicator function that equals 1 if the distance \(d\) matches the discretized distance \(d_j\), and 0 otherwise. $\bm{\lambda}^i_j$ is the $j$-th entry of  $\bm{\lambda}^i$.
The range and granularity of the parameter can be adjusted by setting the value of $d_r$

\subsection{Equilibrium solution}
Existing literature \cite{mckelvey1995quantal} has proved that calculating $p_i^{\mathrm{\sss LV}}$ and $p_i^{\mathrm{\sss TV}}$ from Equation \ref{eq:pq}, \ref{eq:pq2}, \ref{eq:e1} and  \ref{eq:e2} is a fixed-point problem. In this study, we use an Expectation-Maximum (EM) algorithm to iteratively solve the probability values and optimize the parameters. The process is displayed in Algorithm \ref{alg:em}.

\begin{algorithm}[H]
\caption{Solving QRE with Expectation-Maximum}\label{alg:em}
\begin{algorithmic}
\STATE 
\STATE {\textsc{Initialize: }}$\omega_0, \lambda_0, p^{\mathrm{\sss LV}}_0, p_0^{\mathrm{\sss TV}}, \epsilon, t=0$ 
\STATE \textbf{While} $|p_{t+1}^{\mathrm{\sss LV}} - p_{t}^{\mathrm{\sss LV}}| > \epsilon$ or $|p_{t+1}^{\mathrm{\sss TV}} - p_{t}^{\mathrm{\sss TV}}| > \epsilon$ \textbf{do}:
\STATE \hspace{0.5cm} ln$L(\omega_t, \lambda_t) \gets \sum_i \sum_j y_{ij}^{\mathrm{\sss LV}} \times log(p_{j,t}^{\mathrm{\sss LV}}) +  \sum_i \sum_j y_{ij}^{\mathrm{\sss TV}} \times log(p_{j,t}^{\mathrm{\sss TV}})$
\STATE \hspace{0.5cm} $\omega_{t+1}, \lambda_{t+1} \gets \arg\min -\mathrm{ln} L(\omega_t, \lambda_t)$
\STATE \hspace{0.5cm} $p_{i,t+1}^{\mathrm{\sss LV}} \gets \frac{exp\{\lambda_{t+1}^{\mathrm{\sss LV}} E[J^{\mathrm{\sss LV}}(\gamma_i^{\mathrm{\sss LV}})]\}}{\sum_{\gamma_j^{\mathrm{\sss LV}}  \in \Gamma^{\mathrm{\sss LV}}} exp\{\lambda_{t+1}^{\mathrm{\sss LV}} E[J^{\mathrm{\sss LV}}(\gamma_j^{\mathrm{\sss LV}})]\}}$
\STATE \hspace{0.5cm} $p_{i,t+1}^{\mathrm{\sss TV}} \gets \frac{exp\{\lambda_{t+1}^{\mathrm{\sss TV}} E[J^{\mathrm{TV}}(\gamma_i^{\mathrm{TV}})]\}}{\sum_{\gamma_j^{\mathrm{\sss TV}}  \in \Gamma^{\mathrm{\sss TV}}} exp\{\lambda_{t+1}^{\mathrm{\sss TV}} E[J^{\mathrm{\sss TV}}(\gamma_j^{\mathrm{\sss TV}})]\}}$
\STATE \hspace{0.5cm} $t \gets t + 1$
\STATE $\textbf{END}$
\end{algorithmic}
\label{alg1}
\end{algorithm}

In Algorithm \ref{alg:em}, we first initialize the parameters and the probability values. 
While the probability values are not converged, we continue to calculate the log-likelihood function using the current probabilities
$p_{j,t}^{\mathrm{\sss LV}}$, $p_{j,t}^{\mathrm{\sss TV}}$ and labels $y_{ij}^{\mathrm{\sss LV}}$, $y_{ij}^{\mathrm{\sss TV}}$.
$y_{ij}^{\mathrm{\sss LV}} = 1$ ($y_{ij}^{\mathrm{\sss TV}} = 1$) indicates that the player LV (player TV) in the $i$-th row of the data chooses the action $j$ in the ground truth trajectory.
Then, parameters $\omega_{t+1}$ and $\lambda_{t+1}$ are updated by minimizing the negative log-likelihood function.
With the updated parameters, $p_{j,t+1}^{\mathrm{\sss LV}}$, $p_{j,t+1}^{\mathrm{\sss TV}}$ are calculated and fed into the next iteration.

\begin{figure*}[t]
  \centering  \includegraphics[width=\linewidth]{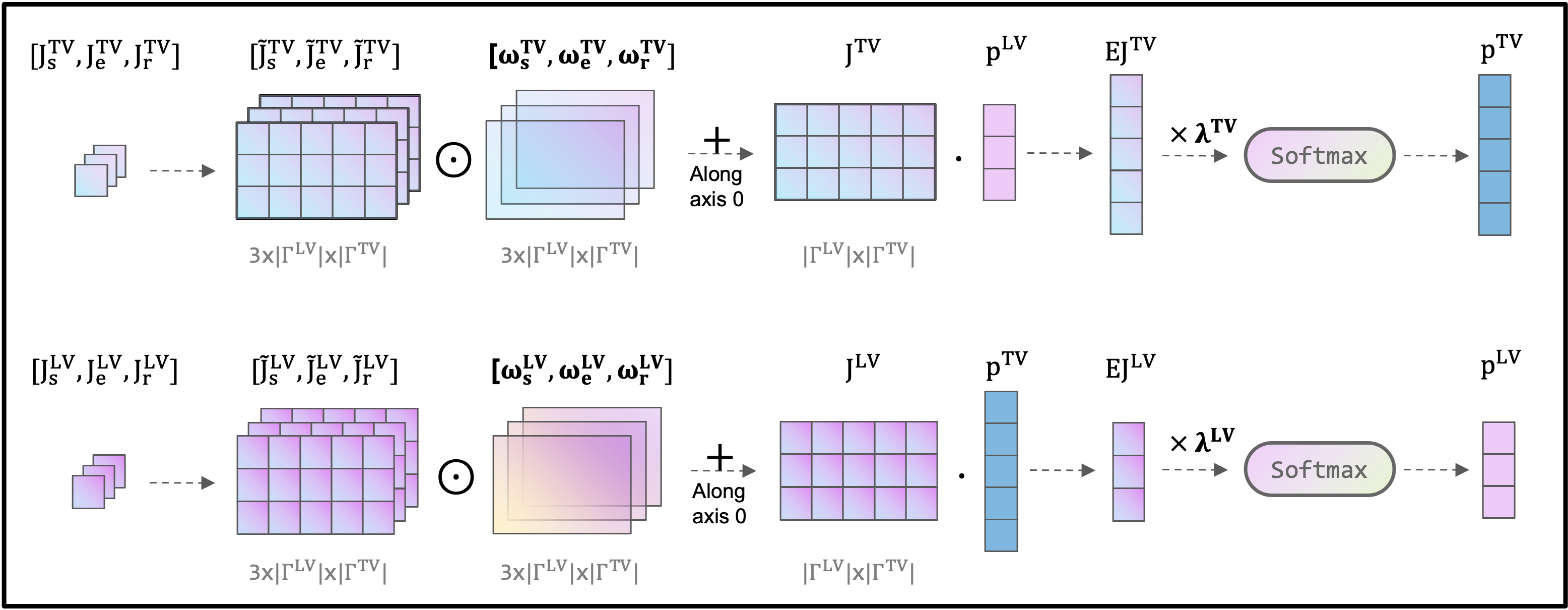}
  \caption{Neural network structure used for parameter learning in EM algorithm}
  \label{fig:network}
\end{figure*}

The core of this algorithm is to calculate the optimal values of $\omega$ and $\lambda$ with maximum likelihood.
In this study, a neural network is utilized to learn optimal values.
As shown in Figure \ref{fig:network}, the input of the network is the payoff vector $[J_s^{\mathrm{\sss LV}}, J_e^{\mathrm{\sss LV}}, J_r^{\mathrm{\sss LV}}]$ and $[J_s^{\mathrm{\sss TV}}, J_e^{\mathrm{\sss TV}}, J_r^{\mathrm{\sss TV}}]$  derived from the current states of the vehicles in the ground truth trajectory data. 
Next, for each action pair $(\gamma_i^{\mathrm{\sss LV}}, \gamma_j^{\mathrm{\sss TV}})$ selected from $(\Gamma^{\mathrm{\sss LV}}, \Gamma^{\mathrm{\sss TV}})$, we assume the left-turning vehicle and the through vehicle make this move at the current time step and calculate the corresponding payoff.
This gives the matrix-form payoff $[\tilde{J} ^i_s, \tilde{J} ^i_e, \tilde{J} ^i_r]$ for the next time step considering possible future actions of both players.
Then, the pointwise product (Hadamard product) of the payoff $[\tilde{J} ^i_s, \tilde{J} ^i_e, \tilde{J} ^i_r]$ and the parameters
$[\omega_s^i, \omega_e^i,\omega_r^i]$ are summed along the index axis to get the overall payoff $J^i$ of vehicle $i$, $i \in \{\mathrm{LV}, \mathrm{TV}\}$.
Then, the payoffs $J^{\mathrm{\sss LV}}$ and $J^{\mathrm{\sss TV}}$ are multiplied by $\lambda^{\mathrm{\sss LV}}$ and $\lambda^{\mathrm{\sss TV}}$ before being sent to a Softmax layer for multiclass classification. 

The fixed-point problem described in Algorithm \ref{alg:em} was first solved using real-world vehicle trajectory data.
Once the probabilities $p_i^{\mathrm{\sss LV}}$ and $p_i^{\mathrm{\sss TV}}$ converged, the learned parameters  were used as inputs to the simulation. In the validation stage, the parameters $\lambda$ and $\omega$ were fixed, and the values of $p_i^{\mathrm{\sss LV}}$ and $p_i^{\mathrm{\sss TV}}$ were evaluated using the same algorithm until the convergence of the two probabilities.

\section{Experimental settings} \label{sec:es}

\subsection{Data}

The decision model was calibrated with CitySim dataset \cite{zheng2022citysim}, a drone-based vehicle trajectory dataset that contains fine-grained vehicle trajectory data and a substantial number of critical safety events.
132 min of vehicle trajectory data with a frequency of 30 Hz from two consecutive signalized intersections (Intersection D and Intersection E) was used.
The position and speed of the vehicles were obtained directly from the data. The longitudinal acceleration of the vehicles was derived from the speed by interpolation.

\begin{figure}[t]
  \centering  \includegraphics[width=\linewidth]{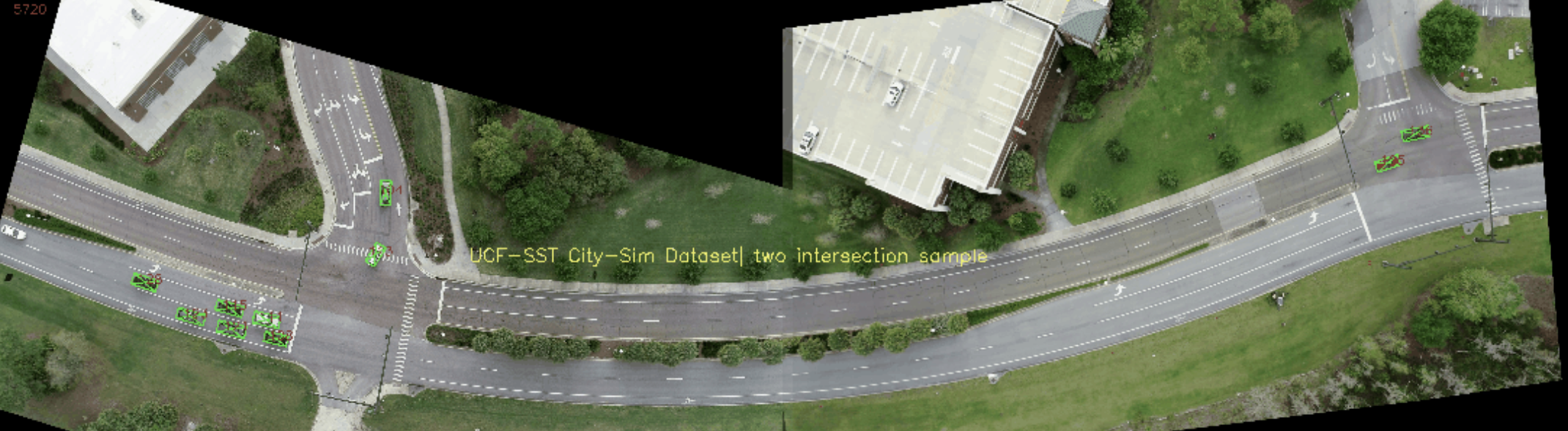}
  \caption{Data collection site}
  \label{fig:site}
\end{figure}

Figure \ref{fig:site} illustrates the data collection site and provides a representation of the intersection geometry. 
The two intersections are located on both ends of a large garage in Florida, USA. 
Both intersections have permitted left-turn signal phases, which lead to potential conflicts between southbound through traffic on the major road and left-turning vehicles heading for the garage.
Combined with the high definition (HD) map and the traffic signal timing information, it is able to extract vehicle unprotected left-turning behavior in the two intersections.

In summary, we collected 819 valid frames containing conflict maneuvers of a left-turning vehicle and a through vehicle. 
70\% of the data was utilized for the model calibration in the training stage, and the rest was used for model validation and testing.



\subsection{Model training}

When solving the QRE with the EM algorithm, the initial payoff weights were set as follows, 
\begin{align*}
    \omega^{\mathrm{\sss LV}}_s = \omega^{\mathrm{\sss TV}}_s = 0.5\\
    \omega^{\mathrm{\sss LV}}_e = \omega^{\mathrm{\sss TV}}_e = 0.3\\
    \omega^{\mathrm{\sss LV}}_r = \omega^{\mathrm{\sss TV}}_r = 0.2
\end{align*}
with reference to the weights of a moderate-style driver from the relevant literature \cite{hang2021cooperative, hang2020human}.
The initial bound rationality parameters were set to 2.0.
Initially, each action has an equal probability of being chosen,
$
    p_0^{\mathrm{\sss LV}} = \begin{bmatrix}
        \frac{1}{3}& \frac{1}{3}& \frac{1}{3}
    \end{bmatrix}^\mathrm{T},
    p_0^{\mathrm{\sss TV}} = \begin{bmatrix}
        \frac{1}{5}&
        \frac{1}{5}&
        \frac{1}{5}&
        \frac{1}{5}&
        \frac{1}{5}
    \end{bmatrix}^\mathrm{T}
$.
Prior to training, we normalized the payoff values to the range [0, 1].
The maximum iteration number of the EM algorithm was set at 20 and the stop threshold $\epsilon$ was set to 0.01.

During learning, a Gaussian kernel is applied to smooth payoff weights, effectively reducing variations and enhancing spatial coherence across different actions. Subsequently, normalization is performed to ensure that, at each position, the weights across the three indices sum to one.
The cross-entropy loss function was used as the optimization objective to measure the discrepancy between predicted probabilities and true labels. The network was optimized using the stochastic gradient descent (SGD) optimizer with a learning rate of 0.1. The batch size was 32. 
The training process in each EM iteration was conducted with a fixed number of epochs set to 30. The parameters learned at the epoch with the highest testing accuracy were selected as the output.

All experiments were carried out on a machine equipped with an NVIDIA GeForce RTX 3060 GPU. We implemented the neural network using PyTorch version 2.0.0.

\subsection{Model validation}
Given a set of calibrated parameters, the model was validated with simulation. 
The simulation platform is established using Python 3.7 and PyQT 5.
In this study, we present a vehicle using a rectangle bounding its geometric contour.
The speed and position of both left-turning vehicle and through vehicle are updated according to the following equations:
\begin{align}
    v_{t+1} &= v_t + \tilde{a}_{t} \Delta t \\
    d_{t+1} &= d_t + v_t \Delta t + \frac{1}{2} \tilde{a}_{t} \Delta t^2
\end{align}
where $\tilde{a}_t$ is the modified longitudinal acceleration at time $t$, and $v_t$ and $d_t$ are the corresponding longitudinal speed and travel distance.
The game model calculates a probability distribution over the action space of LV and TV under QRE. To utilize this unique feature of QRE and also consider the temporal change of bounded rationality in the dynamic decision-making process, we propose a modified longitudinal acceleration $\tilde{a}$ considering drivers' dynamic bounded rationality.
\begin{align}
    \tilde{a}_t^{\mathrm{\sss LV}} = &  \alpha^{\mathrm{\sss LV}} \sum_{\gamma_j^{\mathrm{\sss LV}}  \in \Gamma^{\mathrm{\sss LV}}} \gamma_j^{\mathrm{\sss LV}} \mathbb{I}_{\{p_{j,t}^{\mathrm{\sss LV}} = \max(p_{t}^{\mathrm{\sss LV}})\}} \\ + &(1-\alpha^{\mathrm{\sss LV}})\sum_{\gamma_j^{\mathrm{\sss LV}}  \in \Gamma^{\mathrm{\sss LV}}}\gamma_j^{\mathrm{\sss LV}} p_{j,t}^{\mathrm{\sss LV}} \\
    \tilde{a}_t^{\mathrm{\sss TV}} = & \alpha^{\mathrm{\sss TV}}\sum_{\gamma_j^{\mathrm{\sss TV}}  \in \Gamma^{\mathrm{\sss TV}}} \gamma_j^{\mathrm{\sss TV}}\mathbb{I}_{\{p_{j,t}^{\mathrm{\sss TV}} = \max(p_{t}^{\mathrm{\sss TV}})\}} \\ + & (1-\alpha^{\mathrm{\sss TV}})\sum_{\gamma_j^{\mathrm{\sss TV}}  \in \Gamma^{\mathrm{\sss TV}}} \gamma_j^{\mathrm{\sss TV}}p_{j,t}^{\mathrm{\sss TV}}   
\end{align}
where the first summation term $\sum_{\gamma_j^{i}  \in \Gamma^{i}} \gamma_j^{i} \mathbb{I}_{\{p_{j,t}^{i} = \max(p_{t}^{i})\}}$, $i \in \{\mathrm{LV}, \mathrm{TV}\}$ is the  acceleration rate with the highest corresponding choice probability at time-step $t$. The second summation term $\sum_{\gamma_j^{i}  \in \Gamma^{i}}\gamma_j^{i} p_{j,t}^{i}$, $i \in \{\mathrm{LV}, \mathrm{TV}\}$ is the expectation of vehicles' acceleration over the calculated probability distribution $p_t^i$ in time step $t$.
$\alpha^{i} = e^{-k(d_{conf}^{i} - d_0)} \cdot \mathbb{I}_{\{ d_{conf}^{i} > d_0\}} + 1 \cdot \mathbb{I}_{\{ d_{conf}^{i} \le d_0\}}$ is the dynamic rationality weight related to the distance to conflict point, $i \in \{\mathrm{LV}, \mathrm{TV}\}$, $k=0.1$, $d_0=1.0$ m. 
In this setting, the action with the highest probability will have a higher weight when the vehicle is close to the conflict point and is supposed to be rational.
A suboptimal but reasonable action by calculating the acceleration expectation will tend to be adopted when the vehicle is approaching the conflict point from a long distance.
Future study might adopt a dedicated longitudinal speed controller to consider vehicle dynamic.
The update frequency of the speed and acceleration is 10 Hz. It is assumed that actions can be fulfilled instantaneously.

It is worthwhile to mention that we assume a vehicle plans a path from its origin lane to its target lane before entering the intersection, and follows this pre-planned path to pass through the intersection. In case of conflicts with other vehicles, players only adjust the speed along the path but do not change to other paths. Hence, the action set only consists of vehicle's longitudinal acceleration rate.
This assumption is also adopted in the literature \cite{chen2015cooperative, li2020game}. The pre-planned path is extracted from the ground truth trajectories. We use a cubic polynomial curve to fit the trajectory.
The real-time longitudinal speed and heading angles are derived accordingly based on the analysis formula of the polynomial curve.

In the simulation, we initialize the vehicles based on the positional and velocity data extracted from real-world conflicting trajectories, facilitating a comparative analysis of the simulated scenario against ground truth trajectories. Only the interaction of two vehicles is considered in the simulation.
The decision-making model only accounts for the interaction when both vehicles have not passed the conflict point. When one of the vehicles passes it, the simulation module will set the acceleration of both vehicles to the maximum within the action set before the vehicles reach a maximum speed of 20 m/s. Subsequently, the vehicles maintain a constant speed until they approach the target exit lanes.

The performance of the model was evaluated using simulation completion time (SCT) and fuel consumption. 
Simulation completion time is defined as the duration from the initialization of the simulation to the time instant when both vehicles reach their destination points. 
Fuel consumption is calculated with a vehicle specific power (VSP)-based method \cite{zhao2015development}. The total fuel consumption was assessed for both interacting vehicles.
The safety performance of the decision model is evaluated by the total number of collisions that occurred in  simulations. The speed and acceleration profile will also be displayed and analyzed.

\section{Results} \label{sec:r}

\subsection{Learning results}

\begin{figure}[h]
  \centering  \includegraphics[width=\linewidth]{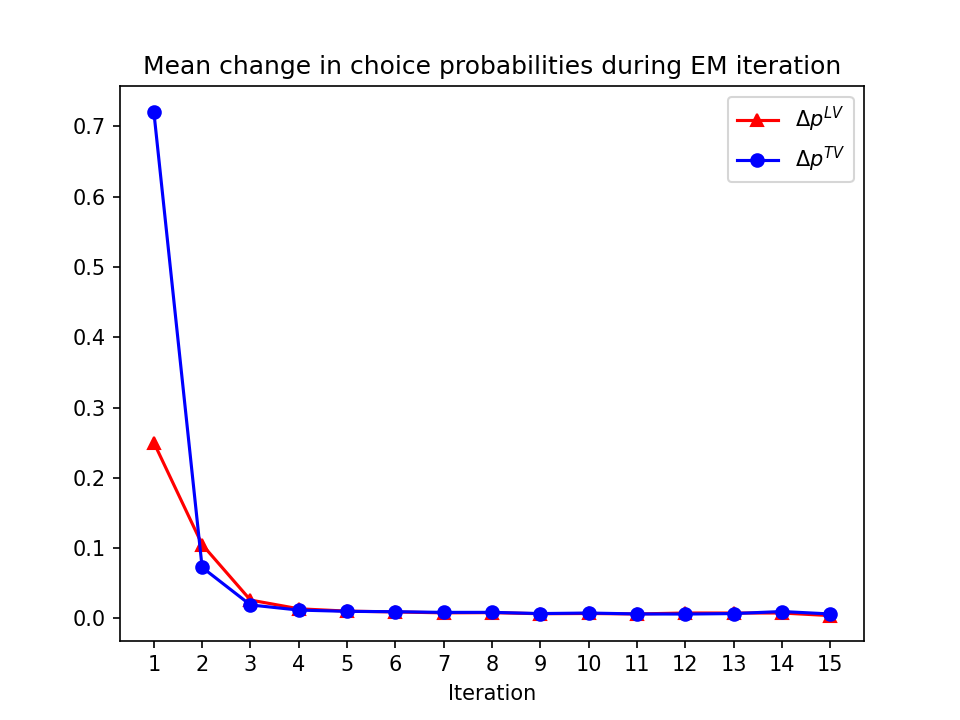}
  \caption{Mean change in probabilities during EM iteration}
  \label{fig:p-diff}
\end{figure}

We plot the mean change of $p^{\mathrm{\sss TV}}$ and $p^{\mathrm{\sss TV}}$ throughout the EM iteration in Figure \ref{fig:p-diff}. The difference in $p^{\mathrm{\sss TV}}$ and $p^{\mathrm{\sss TV}}$ reached the threshold of $\epsilon = 0.01$ in the 15-th iteration. Thus, the convergence of the EM algorithm is confirmed. 

Compared with the ground truth data in the test set, 78.8\% and 78.5\% of our predictions on longitudinal decisions of LV and TV were in line. Note that the prediction task in this paper is a real-time decision prediction on vehicle longitudinal maneuvers. Related research \cite{rahmati2021helping} achieves an accuracy of 82.1\% for a simultaneous game on behavior-level decisions, that is, turning or waiting. In this context, the prediction accuracy of the proposed model is satisfactory for longitudinal decision-making modeling.

\begin{figure*}[t]
  \centering  
  \subfloat[$\omega^{\mathrm{\sss LV}}$]
  {\includegraphics[width=\linewidth]{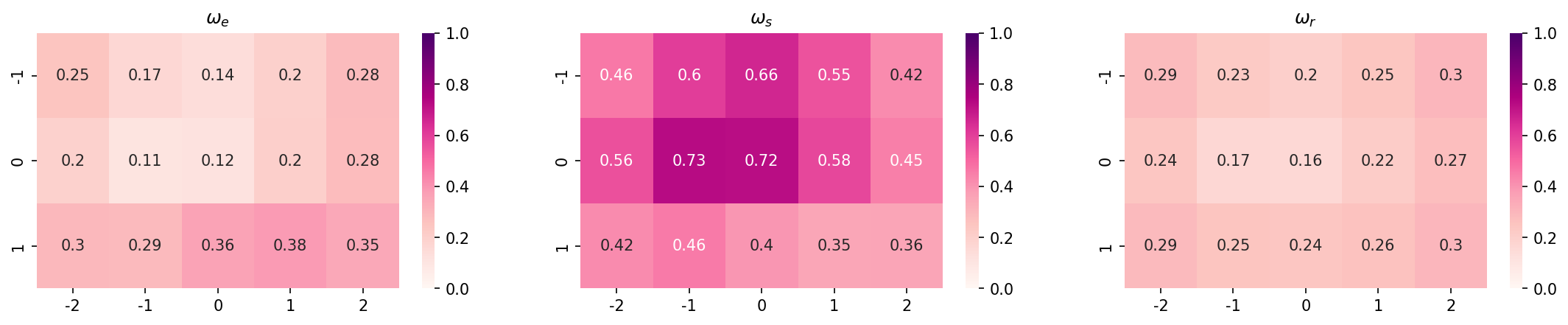}\label{fig:heat-lv}}
\hfil
  \subfloat[$\omega^{\mathrm{\sss TV}}$]
  {\includegraphics[width=\linewidth]{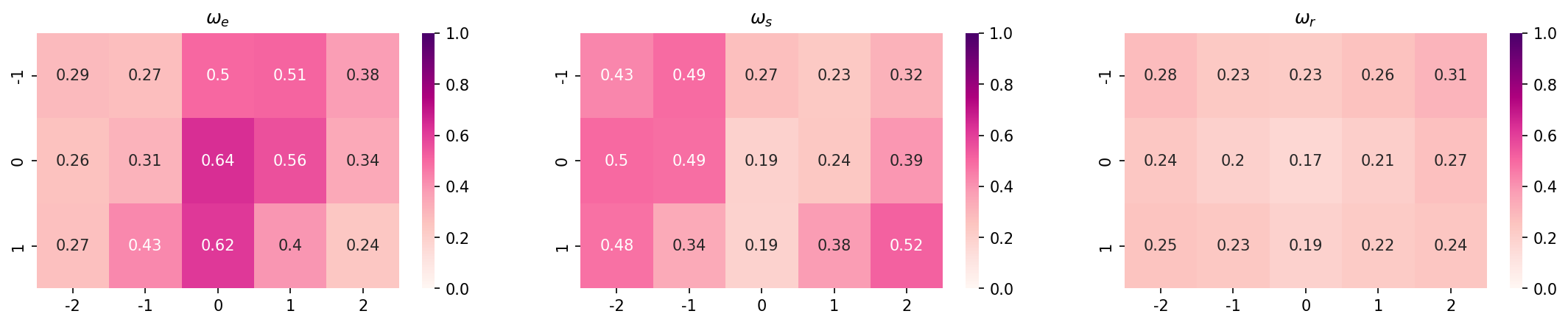}\label{fig:heat-tv}}
  \caption{Heatmap of learned payoff weights} \label{fig:heat}
\end{figure*}

Figure \ref{fig:heat-lv} and Figure \ref{fig:heat-tv} illustrate the learned parameters $\omega^{\mathrm{\sss LV}}$ and $\omega^{\mathrm{\sss TV}}$, respectively, with heatmaps.
To compare among three indexes, we calculate the mean values of $\omega_e$, $\omega_s$ and $\omega_r$ of both LV and TV, as shown in Table \ref{tab:omega-mean}. 
For TV, the payoff weights for driving efficiency and safety received relatively high scores, followed by the weights assigned to traffic rules. In contrast, LV placed a higher priority on safety. These results highlight the general driving styles of TV and LV when making decisions based on the three types of payoff.

In addition, the heatmaps demonstrate the tendency of LV and TV to make different decisions. For each payoff weight matrix, different action pairs receive different weights, with higher weights (deeper colors) indicating a higher decision tendency.
For example, for the heatmap of $\omega_s^{\mathrm{\sss LV}}$, when $\gamma^{\mathrm{\sss TV}} = 0$, the payoff weight for the action -1, 0 and 1 of LV is accordingly 0.66, 0.72 and 0.40. It indicates that when LV takes conservative actions, such as decelerating or keeping speed, it will attach more importance to the safety index. When LV and TV both accelerate, more attention of LV will be drawn to efficiency ($\omega_e^{\mathrm{\sss LV}}(\gamma^{\mathrm{\sss LV}}=1, \gamma^{\mathrm{\sss TV}}=1) = 0.38$), and less to safety.

\begin{figure}[t!]
  \centering  \includegraphics[width=\linewidth]{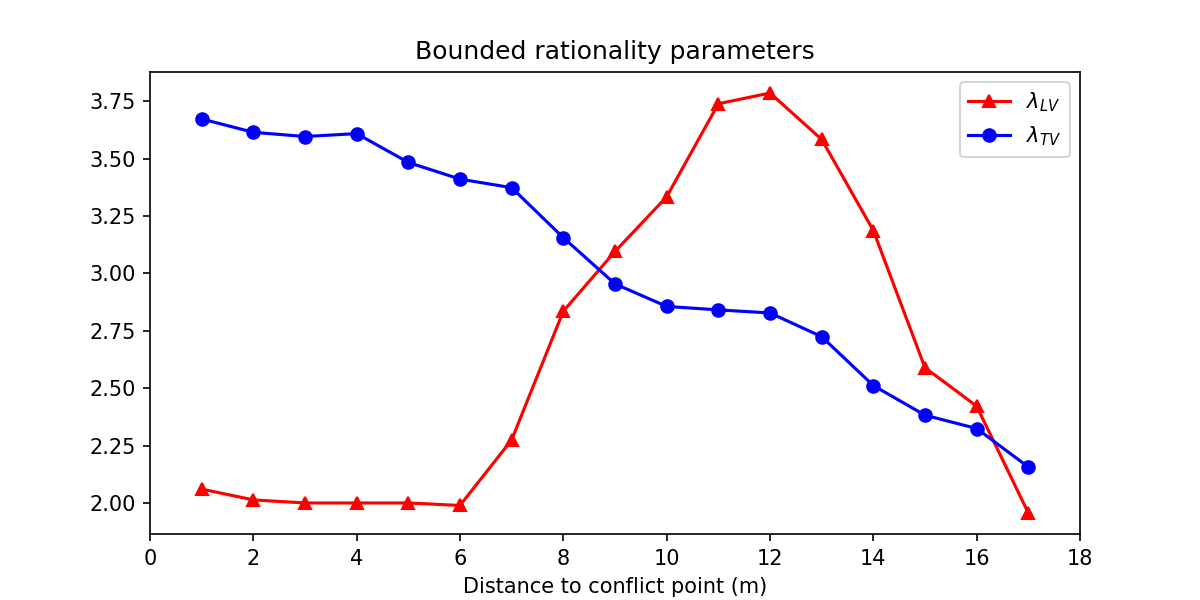}
  \caption{learned interaction-aware bounded rationality parameters.}
  \label{fig:rationality-param}
\end{figure}

EM algorithm also gives the learned bounded rationality parameters, as shown in Figure \ref{fig:rationality-param}. The results indicate that rationality of LV and TV during interaction change differently by distance to conflict point. The rationality of TV increases as it approaches the conflict point. In contrast, the rationality of LV exhibits a nonmonotonic trend with a peak at 12 m to conflict point.
It should be noted that this set of learned bounded rationality parameters is an overall rationality characteristic of interacting agents in the dataset. Individual modeling can be achieved by fine-tuning the overall parameters.

\begin{table}[t]
\caption{Mean values of payoff weights}\label{tab:omega-mean}
\centering
\begin{tabular}{cccc}
\toprule
Player & $\bar \omega_e$ & $\bar \omega_s$ & $ \bar \omega_r$  \\
\midrule
LV & 0.24 & 0.51 & 0.25 \\
TV & 0.40 & 0.36 & 0.24\\
\bottomrule
\end{tabular}
\end{table}



\subsection{Case study}

In this section, we conduct two case studies representing typical vehicle interaction scenarios to evaluate the performance of the proposed decision-making model, denoted by QRE.
The optimal parameters learned from the training stage will be used in the simulation. 
Additionally, to show the superiority of the proposed model, two baseline models are selected for comparison.
The first baseline model is a two-player normal-form game model that uses a pure strategy  NE as the equilibrium output, denoted by "NE". In addition, the game settings are identical between the QRE model and the NE model. 
The second baseline model is a model only with the initial parameter settings of QRE model, denoted by QRE-0. The comparison between QRE and QRE-0 serves as an ablation experiment to show the ability of the learned parameters to capture the players' decision-making tendency.

\subsubsection{Case study 1: potential collision with nearly equal chance}

In this case study, we evaluate a vehicle interaction scenario where two vehicles stand a nearly equal chance to pass through the intersection, i.e., with similar initial speed and initial distance to conflict point.
The initial settings of this case are shown in Table \ref{tab:caseset}.

\begin{table}[t]
\caption{Initial settings of case studies}\label{tab:caseset}
\centering
\begin{tabular}{cccc}
\toprule
 &  & Speed (km/h) & Distance to conflict point (m) \\
\midrule
 Case study 1 & LV & 10.80 & 23.51 \\
 & TV & 10.80 & 22.56\\
Case study 2 & LV & 14.40 &  15.50\\
 & TV & 28.80 & 33.72\\
\bottomrule
\end{tabular}
\end{table}

\begin{figure}[t]
  \centering  \includegraphics[width=\linewidth]{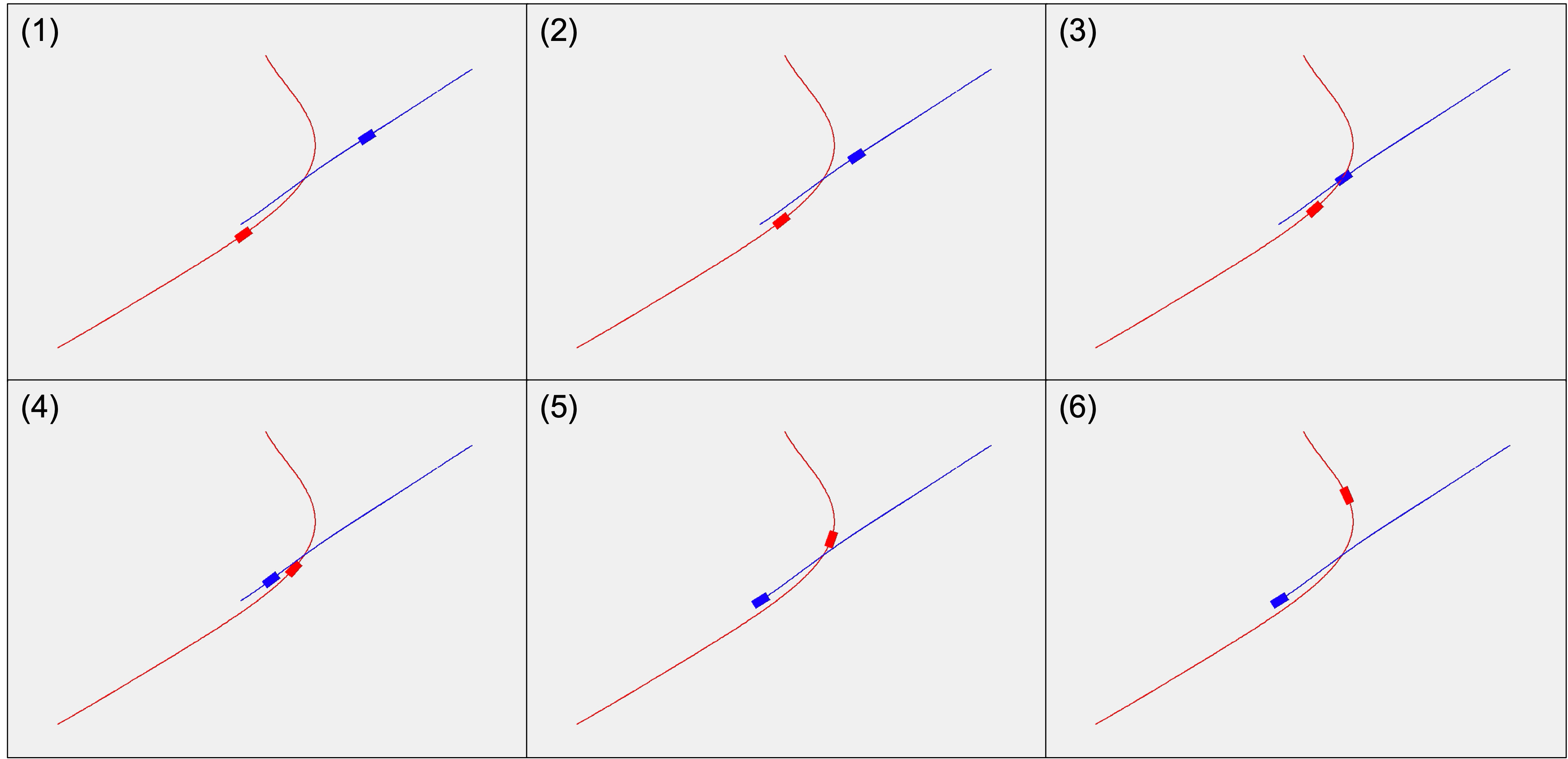}
  \caption{Snapshots of LV and TV positions under QRE in case study 1 (the red rectangle indicates the LV and the blue rectangle indicates the TV)}
  \label{fig:proc}
\end{figure}

We take snapshots every 2 s to describe the positions of both vehicles under QRE, as shown in Figure \ref{fig:proc}. 
It can be seen from the snapshots that no collision occurred between LV and TV. TV passes through the conflict point first and followed by LV.

\begin{figure*}[thb]
  \centering  \includegraphics[width=\linewidth]{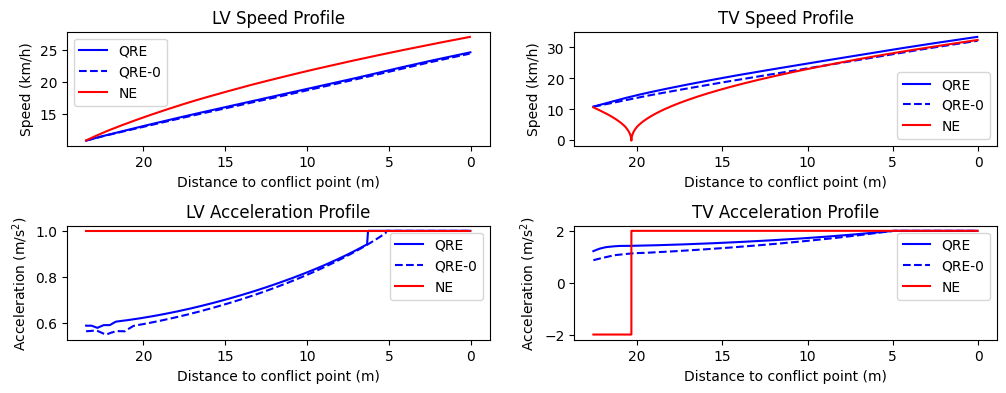}
  \caption{Vehicle speed and acceleration profile of case study 1}
  \label{fig:profile-1}
\end{figure*}

The detailed speed and acceleration profiles of both vehicles under QRE, QRE-0 and NE are shown in Figure \ref{fig:profile-1}, indicated by the blue solid line, blue dash line and red solid line, respectively.
First of all, it can be seen from the speed profiles of LV and TV that vehicles under NE performed an interaction behavior different from that under QRE and QRE-0.
Under NE, TV made a full deceleration and yielded at LV at about 20 m to the conflict point. And LV maintained the maximum acceleration rate all the way before the conflict point.  However, under QRE and QRE-0, TV started at a rather high acceleration rate and passed through the intersection first, while LV gradually increased the acceleration rate and made the turn after TV.
A quantified result of the interaction behavior is presented in Table \ref{tab:case-res}.
The total simulation time of QRE model and QRE-0 model is 10.55 s and 10.62 s respectively, which is lower than that of NE model (11.99 s). The PET of QRE (1.16 s) and QRE-0 (0.97 s) is also sufficiently lower than the PET of NE (4.69 s).

One major reason for the different decision-making outcomes lies in the ability to consider drivers' bounded rationality of the QRE model.
Instead of directly decelerated to yield at LV, TV controlled by QRE and QRE-0 made the "go" decision and demonstrated an increasing acceleration. 
Meanwhile, under QRE, LV started with an acceleration of 0.6 m/s$^2$, but not the max value 1 m/s$^2$ under NE.
This maneuver corresponds to the real-life actions where a left-turning vehicle decides to yield to the through traffic, but still pushes forward to show its intention.
The simulation result confirms that the proposed QRE-based decision-making framework succeeded in modeling this driving behavior, while the NE model failed to do so and led to a different result.


Furthermore, vehicle trajectories under QRE exhibited lower simulation completion time and higher PET compared to those under QRE-0, which means that the interaction behavior given by the proposed model was more efficient and also safe.
This result confirms the usefulness of parameter learning considering drivers' decision tendency to generate more realistic and reliable interaction trajectories.


\begin{table}[t]
\caption{Simulation results of case studies}\label{tab:case-res}
\centering
\begin{tabular}{cccc}
\toprule
 &  & SCP (s) & PET (s)\\
\midrule
Case study 1 & QRE & 10.55 & 1.16 \\
 & QRE-0 & 10.62 & 0.97 \\
 & NE & 11.99 & 4.69 \\
Case study 2 & QRE & 8.61  & 2.17 \\
 & QRE-0 & 8.61 & 2.19\\
 & NE & 9.39 & 3.40\\
\bottomrule
\end{tabular}
\end{table}

\subsubsection{Case study 2: upcoming through vehicle}
In this case study, an interacting scenario is evaluated in which the left-turning vehicle faces a through vehicle that enters the intersection from a further distance but with a higher speed. It is a typical scenario, since the vehicle on the left side of the road with the higher right of way can possibly rush into the intersection and cause potential risks.

\begin{figure*}[thb]
  \centering  \includegraphics[width=\linewidth]{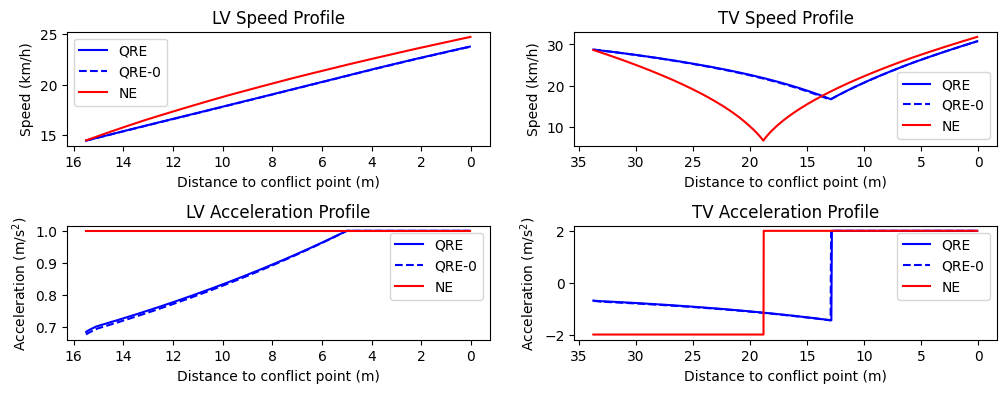}
  \caption{Vehicle speed and acceleration profile of case study 2}
  \label{fig:profile-2}
\end{figure*}

Again, the speed and acceleration profile of LV and TV is shown in Figure \ref{fig:profile-2}.
LV passed through the intersection first under all models. 
TV stopped early with a hash brake before the conflict point under NE. However, under QRE and QRE-0, TV performed a slight deceleration to avoid the collision, consequently making a smoother speed profile.
To give a clear explanation, the QRE model output $p^{\mathrm{\sss LV}}$ and $p^{\mathrm{\sss TV}}$ at the first time-step is given: $p^{\mathrm{\sss LV}} = [0.23, 0.32, 0.45]$, $p^{\mathrm{\sss TV}} = [0.29, 0.21, 0.15, 0.18, 0.16]$. 
Although the actions with the highest probabilities are $\gamma^{\mathrm{\sss LV}} = 1$ and $\gamma^{\mathrm{\sss TV}} = -2$, the probabilities were nearly uniformly distributed and there is no dominant action.
Hence, the QRE model yields smooth movements for both vehicles, since it considers the bounded rationality of vehicles, as well as the dynamic change of bounded rationality during vehicles approaching the conflict point.
Subsequently, the QRE model and QRE-0 model achieved significantly higher efficiency, as shown in Table \ref{tab:case-res}.

\subsection{Overall performance}
To obtain a more comprehensive understanding of the performance of the proposed model in different settings, we run the simulation 1,000 times with random initial positions and longitudinal speeds of both vehicles. 
The initial speed is sampled from a uniform distribution ranging from 10 km/h to 36 km/h.
The initial position is represented by the distance to the conflict point, which is sampled from a uniform distribution ranging from 10 to 40 m.
Again, we compare the proposed model with two baseline decision-making models, QRE-0 and NE.


In the 1000-times simulation, 19, 15 and 23 collisions in total had occurred under QRE, QRE-0 and NE, respectively.
The average simulation completion time for QRE, QRE-0 and NE was 9.11 s, 9.17 s, and 10.20 s.
The average fuel consumption for QRE, QRE-0 and NE was 37.16 mL, 37.22 mL, 42.98 mL.
In rough terms, the proposed QRE model can handle the vehicle interaction problem more efficiently and smoothly than QRE-0 and NE, while maintaining driving safety.

\begin{table*}[tbp]
\caption{Mean values of simulation completion time and total fuel consumption for different initial RTTCs
}\label{tab:overall}
\centering
\begin{tabular}{lllllllr}
\toprule
RTTC (s) & \multicolumn{3}{c}{Simulation completion time (s)} & \multicolumn{3}{c}{Fuel consumpation (mL)} & $^\#$Samples \\
\cline{2-7}
 & QRE          & QRE-0  & NE        & QRE & QRE-0 & NE &  \\
\midrule
0                & $\textbf{8.77}^{***}$  & 8.84  & 9.48  & $34.63^{***}$          & \textbf{34.50} & 39.58 & 158   \\
1                & $\textbf{8.64}^{***}$  & 8.67  & 9.44  & $\textbf{35.50}^{***}$ & 35.52          & 40.40 & 274   \\
2                & $\textbf{8.97}^{***}$  & 9.02  & 10.13 & $\textbf{36.68}^{***}$ & 36.78          & 42.89 & 201   \\
3                & $\textbf{9.15}^{***}$  & 9.21  & 10.47 & $\textbf{37.43}^{***}$ & 37.57          & 44.07 & 135   \\
4                & $\textbf{9.78}^{***}$  & 9.84  & 11.28 & $\textbf{39.89}^{***}$ & 40.05          & 46.72 & 78    \\
5                & $\textbf{10.06}^{***}$ & 10.14 & 11.49 & $\textbf{40.91}^{***}$ & 41.15          & 47.51 & 65    \\
6                & $\textbf{9.96}^{***}$  & 10.04 & 11.41 & $\textbf{41.59}^{***}$& 41.79          & 48.17 & 41    \\
7                & $\textbf{10.24}^{**}$ & 10.34 & 11.74 & $\textbf{41.96}^{***}$ & 42.23          & 49.25 & 24    \\
8                & \textbf{10.47} & 10.55 & 11.89 & $\textbf{43.76}^{*}$ & 43.94          & 49.86 & 12    \\
9                & \textbf{10.76} & 10.83 & 12.27 & \textbf{44.26} & 44.43          & 50.23 & 8     \\
$\ge$10 & \textbf{9.74}  & 9.84  & 11.12 & \textbf{42.02} & 42.25          & 48.28 & 4     \\
All cases          & \textbf{9.11}  & 9.17  & 10.20 & \textbf{37.16} & 37.22          & 42.98 & 1000 \\
\bottomrule
\end{tabular}
\end{table*}

To further illustrate the performance of the three models across various scenarios, we calculated the mean values of simulation completion time and fuel consumption for different initial RTTCs (rounded to integers), as presented in Table \ref{tab:overall}. Additionally, we assess the statistical significance of the QRE model's performance compared to the NE model. Significance levels are marked as follows: '*' for p $<$ 0.05, '**' for p $<$ 0.01, and '***' for p $<$ 0.001.
The QRE model achieved the lowest simulation completion time across all scenarios, indicating its robustness in resolving vehicle conflicts efficiently and smoothly. Notably, in urgent and safety-critical interactions, which predominate in the simulation, the QRE model demonstrated statistically significant improvements over NE model. For example, the QRE model reduced the simulation completion time by 10.55\% in scenarios where RTTC was lower than 5 s, compared to the NE model. Our QRE model also achieves a significant reduction in total fuel consumption, decreasing it by 13.54\% across all cases.

\section{Conclusions} \label{sec:c}

In this study, a game theory-based vehicle unprotected left-turn decision-making model considering drivers' bounded rationality is proposed. The model is realized through a quantal response equilibrium (QRE) defined in a two-player normal-form game. The model parameters are optimized by an EM algorithm with a subtle neural network trained by high-precision microscopic vehicle trajectory data.
The model is validated by two typical simulation case studies and repeated  experiments with random initialization. 
Through simulation experiments, the superiority of the proposed model in accurately capturing drivers' bounded rationality is confirmed. Compared to two baseline models, the QRE model showed a 1.5\% increase in efficiency in interacting scenarios. Under the proposed model, vehicles can also behave socially and show intentions. The learned parameters considering drivers' decision tendency and interaction-aware bounded rationality parameters are also proved to be valuable.

The theoretical analysis of the non-cooperative game with QRE sheds some insightful lights on understanding vehicle decision-making and collision avoidance behavior with bounded rationality in left-turn scenario. The results of this study can be directly incorporated into autonomous driving systems and help to accelerate the realization of CAV technology. This could substantially improve the efficiency of vehicle interaction at intersections and, at the same time, facilitate greater safety by reducing randomness and human error.

One future direction of research may include modeling of drivers' bounded rationality in multi-vehicle interacting scenarios. Additionally, exploring drivers' distinct driving style and rationality warrants further investigation.

\section*{Acknowledgement(s)}
This research was supported by grants from National Key Research and Development Program of China (2022YFB2503200), Tsinghua University-Mercedes Benz Joint Institute for Sustainable Mobility.

\bibliographystyle{IEEEtran}
\bibliography{ref}  

\begin{thebibliography}{10}
\providecommand{\url}[1]{#1}
\csname url@samestyle\endcsname
\providecommand{\newblock}{\relax}
\providecommand{\bibinfo}[2]{#2}
\providecommand{\BIBentrySTDinterwordspacing}{\spaceskip=0pt\relax}
\providecommand{\BIBentryALTinterwordstretchfactor}{4}
\providecommand{\BIBentryALTinterwordspacing}{\spaceskip=\fontdimen2\font plus
\BIBentryALTinterwordstretchfactor\fontdimen3\font minus \fontdimen4\font\relax}
\providecommand{\BIBforeignlanguage}[2]{{%
\expandafter\ifx\csname l@#1\endcsname\relax
\typeout{** WARNING: IEEEtran.bst: No hyphenation pattern has been}%
\typeout{** loaded for the language `#1'. Using the pattern for}%
\typeout{** the default language instead.}%
\else
\language=\csname l@#1\endcsname
\fi
#2}}
\providecommand{\BIBdecl}{\relax}
\BIBdecl

\bibitem{li2020game}
N.~Li, Y.~Yao, I.~Kolmanovsky, E.~Atkins, and A.~R. Girard, ``Game-theoretic modeling of multi-vehicle interactions at uncontrolled intersections,'' \emph{IEEE Transactions on Intelligent Transportation Systems}, vol.~23, no.~2, pp. 1428--1442, 2020.

\bibitem{choi2010crash}
E.-H. Choi, ``Crash factors in intersection-related crashes: An on-scene perspective,'' National Highway Traffic Safety Administration, Tech. Rep. DOT HS 811 366, 2010.

\bibitem{wang2019enabling}
Y.~Wang, Y.~Ren, S.~Elliott, and W.~Zhang, ``Enabling courteous vehicle interactions through game-based and dynamics-aware intent inference,'' \emph{IEEE Transactions on Intelligent Vehicles}, vol.~5, no.~2, pp. 217--228, 2019.

\bibitem{ali2019game}
Y.~Ali, Z.~Zheng, M.~M. Haque, and M.~Wang, ``A game theory-based approach for modelling mandatory lane-changing behaviour in a connected environment,'' \emph{Transportation Research Part C: Emerging Technologies}, vol. 106, pp. 220--242, 2019.

\bibitem{lopez2022game}
V.~G. Lopez, F.~L. Lewis, M.~Liu, Y.~Wan, S.~Nageshrao, and D.~Filev, ``Game-theoretic lane-changing decision making and payoff learning for autonomous vehicles,'' \emph{IEEE Transactions on Vehicular Technology}, vol.~71, no.~4, pp. 3609--3620, 2022.

\bibitem{wang2021competitive}
H.~Wang, Q.~Meng, S.~Chen, and X.~Zhang, ``Competitive and cooperative behaviour analysis of connected and autonomous vehicles across unsignalised intersections: A game-theoretic approach,'' \emph{Transportation Research Part B: Methodological}, vol. 149, pp. 322--346, 2021.

\bibitem{rahmati2017towards}
Y.~Rahmati and A.~Talebpour, ``Towards a collaborative connected, automated driving environment: A game theory based decision framework for unprotected left turn maneuvers,'' in \emph{2017 IEEE Intelligent Vehicles Symposium (IV)}.\hskip 1em plus 0.5em minus 0.4em\relax IEEE, 2017, pp. 1316--1321.

\bibitem{rahmati2021helping}
Y.~Rahmati, M.~K. Hosseini, and A.~Talebpour, ``Helping automated vehicles with left-turn maneuvers: A game theory-based decision framework for conflicting maneuvers at intersections,'' \emph{IEEE Transactions on Intelligent Transportation Systems}, vol.~23, no.~8, pp. 11\,877--11\,890, 2021.

\bibitem{qin2024game}
Z.~Qin, A.~Ji, Z.~Sun, G.~Wu, P.~Hao, and X.~Liao, ``Game theoretic application to intersection management: A literature review,'' \emph{IEEE Transactions on Intelligent Vehicles}, 2024.

\bibitem{sivak2002common}
M.~Sivak, ``How common sense fails us on the road: contribution of bounded rationality to the annual worldwide toll of one million traffic fatalities,'' \emph{Transportation Research Part F: Traffic Psychology and Behaviour}, vol.~5, no.~4, pp. 259--269, 2002.

\bibitem{ahmane2013modeling}
M.~Ahmane, A.~Abbas-Turki, F.~Perronnet, J.~Wu, A.~El~Moudni, J.~Buisson, and R.~Zeo, ``Modeling and controlling an isolated urban intersection based on cooperative vehicles,'' \emph{Transportation Research Part C: Emerging Technologies}, vol.~28, pp. 44--62, 2013.

\bibitem{de2013autonomous}
G.~R. de~Campos, P.~Falcone, and J.~Sj{\"o}berg, ``Autonomous cooperative driving: A velocity-based negotiation approach for intersection crossing,'' in \emph{16th International IEEE Conference on Intelligent Transportation Systems (ITSC 2013)}.\hskip 1em plus 0.5em minus 0.4em\relax IEEE, 2013, pp. 1456--1461.

\bibitem{lee2012development}
J.~Lee and B.~Park, ``Development and evaluation of a cooperative vehicle intersection control algorithm under the connected vehicles environment,'' \emph{IEEE Transactions on Intelligent Transportation Systems}, vol.~13, no.~1, pp. 81--90, 2012.

\bibitem{chen2021rhythmic}
X.~Chen, M.~Li, X.~Lin, Y.~Yin, and F.~He, ``Rhythmic control of automated traffic—part i: Concept and properties at isolated intersections,'' \emph{Transportation Science}, vol.~55, no.~5, pp. 969--987, 2021.

\bibitem{lin2021rhythmic}
X.~Lin, M.~Li, Z.-J.~M. Shen, Y.~Yin, and F.~He, ``Rhythmic control of automated traffic—part ii: Grid network rhythm and online routing,'' \emph{Transportation Science}, vol.~55, no.~5, pp. 988--1009, 2021.

\bibitem{liu2024integrated}
Q.~Liu, K.~Zhang, M.~Li, X.~Chen, X.~Lin, and S.~Li, ``Integrated optimization of traffic signal timings and vehicle trajectories considering mandatory lane-changing at isolated intersections,'' \emph{Transportation Research Part C: Emerging Technologies}, vol. 163, p. 104614, 2024.

\bibitem{schwarting2017safe}
W.~Schwarting, J.~Alonso-Mora, L.~Paull, S.~Karaman, and D.~Rus, ``Safe nonlinear trajectory generation for parallel autonomy with a dynamic vehicle model,'' \emph{IEEE Transactions on Intelligent Transportation Systems}, vol.~19, no.~9, pp. 2994--3008, 2017.

\bibitem{basil2023evaluation}
N.~Basil, M.~Alqaysi, M.~Deveci, A.~Albahri, O.~Albahri, and A.~Alamoodi, ``Evaluation of autonomous underwater vehicle motion trajectory optimization algorithms,'' \emph{Knowledge-Based Systems}, vol. 276, p. 110722, 2023.

\bibitem{you2019advanced}
C.~You, J.~Lu, D.~Filev, and P.~Tsiotras, ``Advanced planning for autonomous vehicles using reinforcement learning and deep inverse reinforcement learning,'' \emph{Robotics and Autonomous Systems}, vol. 114, pp. 1--18, 2019.

\bibitem{liu2023towards}
J.~Liu, D.~Zhou, P.~Hang, Y.~Ni, and J.~Sun, ``Towards socially responsive autonomous vehicles: A reinforcement learning framework with driving priors and coordination awareness,'' \emph{IEEE Transactions on Intelligent Vehicles}, 2023.

\bibitem{liu2022three}
M.~Liu, Y.~Wan, F.~L. Lewis, S.~Nageshrao, and D.~Filev, ``A three-level game-theoretic decision-making framework for autonomous vehicles,'' \emph{IEEE Transactions on Intelligent Transportation Systems}, vol.~23, no.~11, pp. 20\,298--20\,308, 2022.

\bibitem{fang2024cooperative}
S.~Fang, P.~Hang, C.~Wei, Y.~Xing, and J.~Sun, ``Cooperative driving of connected autonomous vehicles in heterogeneous mixed traffic: A game theoretic approach,'' \emph{IEEE Transactions on Intelligent Vehicles}, 2024.

\bibitem{niels2024optimization}
T.~Niels, K.~Bogenberger, M.~Papageorgiou, and I.~Papamichail, ``Optimization-based intersection control for connected automated vehicles and pedestrians,'' \emph{Transportation research record}, vol. 2678, no.~2, pp. 135--152, 2024.

\bibitem{li2018humanlike}
L.~Li, K.~Ota, and M.~Dong, ``Humanlike driving: Empirical decision-making system for autonomous vehicles,'' \emph{IEEE Transactions on Vehicular Technology}, vol.~67, no.~8, pp. 6814--6823, 2018.

\bibitem{wang2024improving}
J.~Wang, Z.~Jiang, and Y.~V. Pant, ``Improving safety in mixed traffic: A learning-based model predictive control for autonomous and human-driven vehicle platooning,'' \emph{Knowledge-Based Systems}, p. 111673, 2024.

\bibitem{hou2024merging}
X.~Hou, M.~Gan, W.~Wu, C.~Wang, Y.~Ji, and S.~Zhao, ``Merging planning in dense traffic scenarios using interactive safe reinforcement learning,'' \emph{Knowledge-Based Systems}, p. 111548, 2024.

\bibitem{lu2023game}
X.~Lu, H.~Zhao, C.~Li, B.~Gao, and H.~Chen, ``A game-theoretic approach on conflict resolution of autonomous vehicles at unsignalized intersections,'' \emph{IEEE Transactions on Intelligent Transportation Systems}, 2023.

\bibitem{jing2019cooperative}
S.~Jing, F.~Hui, X.~Zhao, J.~Rios-Torres, and A.~J. Khattak, ``Cooperative game approach to optimal merging sequence and on-ramp merging control of connected and automated vehicles,'' \emph{IEEE Transactions on Intelligent Transportation Systems}, vol.~20, no.~11, pp. 4234--4244, 2019.

\bibitem{mandiau2008behaviour}
R.~Mandiau, A.~Champion, J.-M. Auberlet, S.~Espi{\'e}, and C.~Kolski, ``Behaviour based on decision matrices for a coordination between agents in a urban traffic simulation,'' \emph{Applied Intelligence}, vol.~28, no.~2, p. 121, 2008.

\bibitem{sadigh2016planning}
D.~Sadigh, S.~Sastry, S.~A. Seshia, and A.~D. Dragan, ``Planning for autonomous cars that leverage effects on human actions.'' in \emph{Robotics: Science and systems}, vol.~2.\hskip 1em plus 0.5em minus 0.4em\relax Ann Arbor, MI, USA, 2016, pp. 1--9.

\bibitem{hang2020human}
P.~Hang, C.~Lv, Y.~Xing, C.~Huang, and Z.~Hu, ``Human-like decision making for autonomous driving: A noncooperative game theoretic approach,'' \emph{IEEE Transactions on Intelligent Transportation Systems}, vol.~22, no.~4, pp. 2076--2087, 2020.

\bibitem{jia2023interactive}
S.~Jia, Y.~Zhang, X.~Li, X.~Na, Y.~Wang, B.~Gao, B.~Zhu, and R.~Yu, ``Interactive decision-making with switchable game modes for automated vehicles at intersections,'' \emph{IEEE Transactions on Intelligent Transportation Systems}, 2023.

\bibitem{yan2023multi}
Y.~Yan, L.~Peng, T.~Shen, J.~Wang, D.~Pi, D.~Cao, and G.~Yin, ``A multi-vehicle game-theoretic framework for decision making and planning of autonomous vehicles in mixed traffic,'' \emph{IEEE Transactions on Intelligent Vehicles}, 2023.

\bibitem{yuan2021deep}
M.~Yuan, J.~Shan, and K.~Mi, ``Deep reinforcement learning based game-theoretic decision-making for autonomous vehicles,'' \emph{IEEE Robotics and Automation Letters}, vol.~7, no.~2, pp. 818--825, 2021.

\bibitem{shen2023analysis}
Z.~Shen, S.~Li, Y.~Liu, and X.~Tang, ``Analysis of driving behavior in unprotected left turns for autonomous vehicles using ensemble deep clustering,'' \emph{IEEE Transactions on Intelligent Vehicles}, 2023.

\bibitem{zhou2019autonomous}
D.~Zhou, Z.~Ma, and J.~Sun, ``Autonomous vehicles’ turning motion planning for conflict areas at mixed-flow intersections,'' \emph{IEEE Transactions on Intelligent Vehicles}, vol.~5, no.~2, pp. 204--216, 2019.

\bibitem{zhou2022autonomous}
D.~Zhou, Z.~Ma, X.~Zhang, and J.~Sun, ``Autonomous vehicles’ intended cooperative motion planning for unprotected turning at intersections,'' \emph{IET Intelligent Transport Systems}, vol.~16, no.~8, pp. 1058--1073, 2022.

\bibitem{zhao2023unprotected}
J.~Zhao, V.~L. Knoop, J.~Sun, Z.~Ma, and M.~Wang, ``Unprotected left-turn behavior model capturing path variations at intersections,'' \emph{IEEE Transactions on Intelligent Transportation Systems}, 2023.

\bibitem{liu2023teaching}
J.~Liu, X.~Qi, Y.~Ni, J.~Sun, and P.~Hang, ``Teaching autonomous vehicles to express interaction intent during unprotected left turns: A human-driving-prior-based trajectory planning approach,'' \emph{arXiv preprint arXiv:2307.15950}, 2023.

\bibitem{wang2022modeling}
B.~Wang, Z.~Li, S.~Wang, M.~Li, and A.~Ji, ``Modeling bounded rationality in discretionary lane change with the quantal response equilibrium of game theory,'' \emph{Transportation Research Part B: Methodological}, vol. 164, pp. 145--161, 2022.

\bibitem{mckelvey1995quantal}
R.~D. McKelvey and T.~R. Palfrey, ``Quantal response equilibria for normal form games,'' \emph{Games and Economic Behavior}, vol.~10, no.~1, pp. 6--38, 1995.

\bibitem{alexiadis2004next}
V.~Alexiadis, J.~Colyar, J.~Halkias, R.~Hranac, and G.~McHale, ``The next generation simulation program,'' \emph{Institute of Transportation Engineers. ITE Journal}, vol.~74, no.~8, p.~22, 2004.

\bibitem{talebpour2015modeling}
A.~Talebpour, H.~S. Mahmassani, and S.~H. Hamdar, ``Modeling lane-changing behavior in a connected environment: A game theory approach,'' \emph{Transportation Research Procedia}, vol.~7, pp. 420--440, 2015.

\bibitem{chen2023game}
W.~Chen, G.~Ren, Q.~Cao, J.~Song, Y.~Liu, and C.~Dong, ``A game-theory-based approach to modeling lane-changing interactions on highway on-ramps: Considering the bounded rationality of drivers,'' \emph{Mathematics}, vol.~11, no.~2, p. 402, 2023.

\bibitem{chan2006characterization}
C.-Y. Chan, ``Characterization of driving behaviors based on field observation of intersection left-turn across-path scenarios,'' \emph{IEEE Transactions on Intelligent Transportation Systems}, vol.~7, no.~3, pp. 322--331, 2006.

\bibitem{chen2017surrogate}
P.~Chen, W.~Zeng, G.~Yu, Y.~Wang \emph{et~al.}, ``Surrogate safety analysis of pedestrian-vehicle conflict at intersections using unmanned aerial vehicle videos,'' \emph{Journal of Advanced Transportation}, vol. 2017, 2017.

\bibitem{zheng2022citysim}
O.~Zheng, M.~Abdel-Aty, L.~Yue, A.~Abdelraouf, Z.~Wang, and N.~Mahmoud, ``Citysim: A drone-based vehicle trajectory dataset for safety-oriented research and digital twins,'' \emph{Transportation Research Record}, p. 03611981231185768, 2022.

\bibitem{hang2021cooperative}
P.~Hang, C.~Lv, C.~Huang, Y.~Xing, and Z.~Hu, ``Cooperative decision making of connected automated vehicles at multi-lane merging zone: A coalitional game approach,'' \emph{IEEE Transactions on Intelligent Transportation Systems}, vol.~23, no.~4, pp. 3829--3841, 2021.

\bibitem{chen2015cooperative}
L.~Chen and C.~Englund, ``Cooperative intersection management: A survey,'' \emph{IEEE Transactions on Intelligent Transportation Systems}, vol.~17, no.~2, pp. 570--586, 2015.

\bibitem{zhao2015development}
X.~Zhao, Y.~Wu, J.~Rong, and Y.~Zhang, ``Development of a driving simulator based eco-driving support system,'' \emph{Transportation Research Part C: Emerging Technologies}, vol.~58, pp. 631--641, 2015.

\end{thebibliography}

\newpage

\vspace{11pt}

\begin{IEEEbiography}
[{\includegraphics[width=1in,height=1.25in,clip]{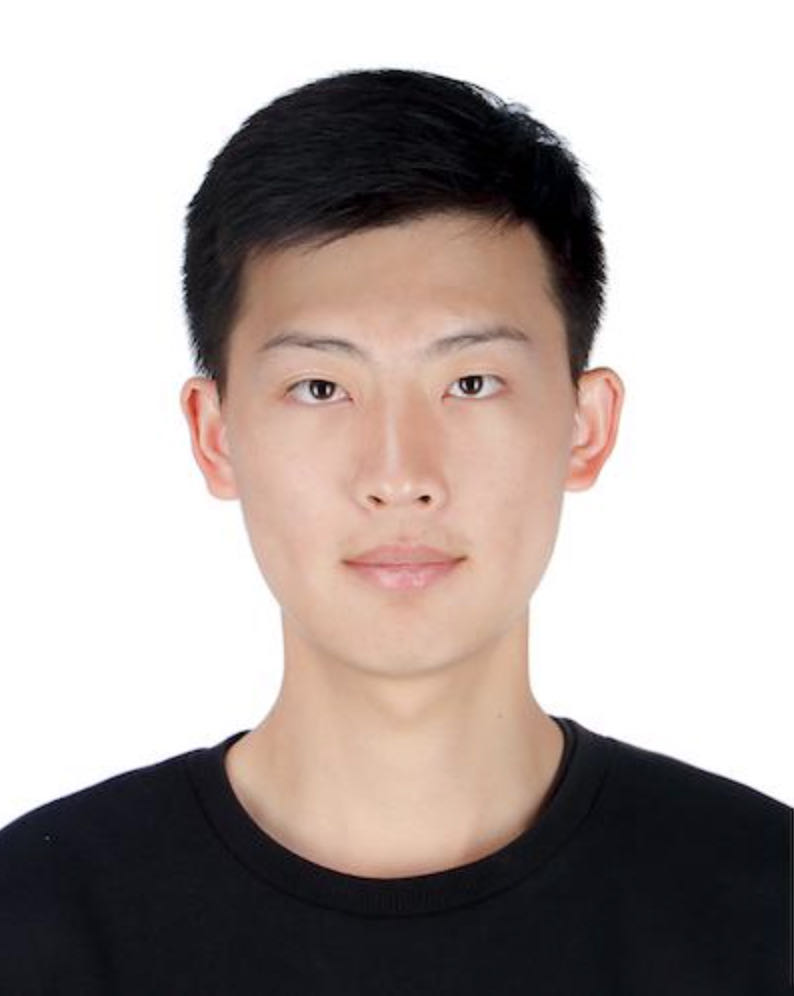}}]{Yuansheng Lian}
received the B.S. degree in civil engineering from Tsinghua University in 2022, where he is currently pursuing the Ph.D. degree with the Department of Civil Engineering. His research interests include intelligent transportation systems and intelligent vehicles.
\end{IEEEbiography}

\begin{IEEEbiography}
[{\includegraphics[width=1in,height=1.25in,clip]{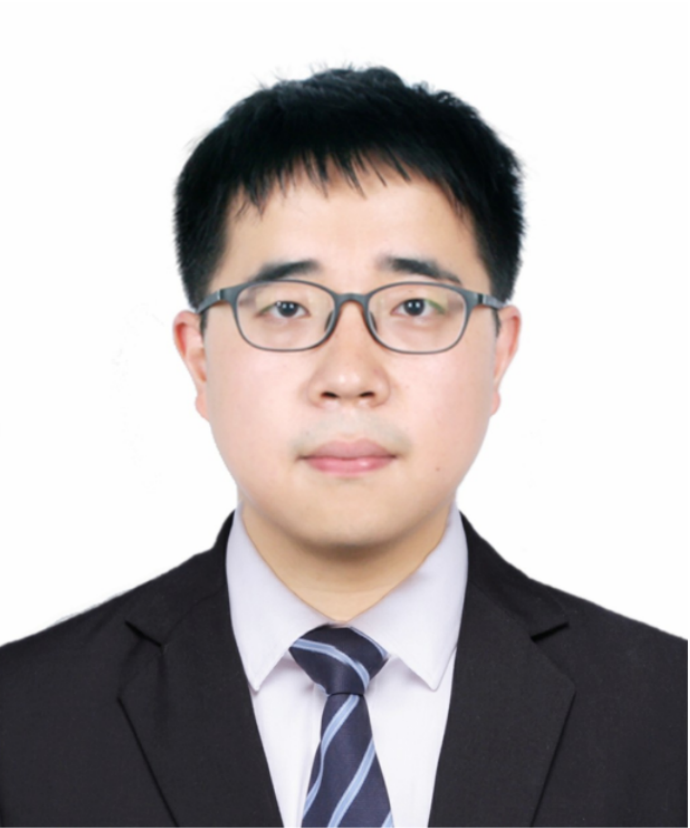}}]{Ke Zhang} received the Ph.D. degree in civil engineering (transportation) from Tsinghua University, Beijing, in 2023. He is currently a postdoctor in Tsinghua University. His research interests include intelligent transportation systems and reinforcement learning. He has won the Outstanding Ph.D. Dissertation Award of Tsinghua University and the Transportation Research Board (TRB) AED50 Best Doctoral Dissertation Award.
\end{IEEEbiography}

\begin{IEEEbiography}
[{\includegraphics[width=1in,height=1.2in,clip]{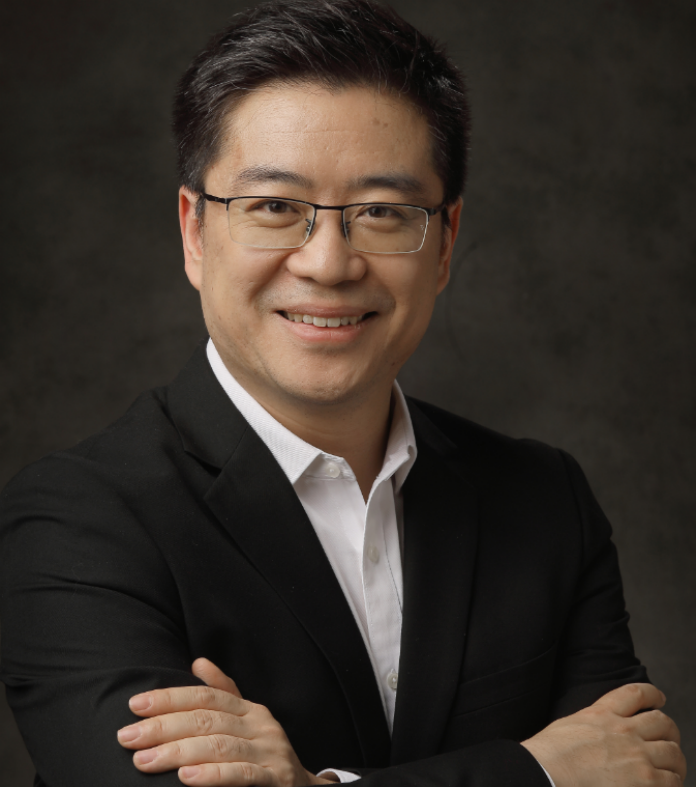}}]{Meng Li} received the B.S. degree in structure engineering from Tsinghua University, Beijing, China, in 2001, the M.S. degree in transportation science and supply chain management and the M.E. degree in transportation engineering from the University of California at Berkeley, Berkeley, USA, in 2003 and 2004, respectively, and the Ph.D. degree in transportation engineering from Nagoya University, Japan, in 2010. He is currently a Professor with the Department of Civil Engineering, Tsinghua University, working in the fields of transportation big data analysis, intelligent transportation systems, and automated transportation. He is the General Scientific Committee Member and the Topic Area Manager (TAM) of the World Conference on Transport Research Society (WCTRS) and the Committee Member of the Sub-Committee of the Transportation Research Board (TRB), USA. He serves as the Editorial Board Member for the Journal of Intelligent Transportation Systems; and an Associate Editor for the Asia–Pacific Journal of Operational Research, IEEE Open Journal of Intelligent Transportation Systems, and Smart and Resilient Transportation.
\end{IEEEbiography}

\begin{IEEEbiography}
[{\includegraphics[width=1in,height=1.4in,clip]{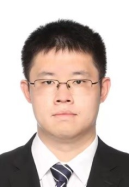}}]{Shen Li} received the Ph.D. degree at the University of Wisconsin–Madison in 2018. He is currently a research associate at Tsinghua University. His research interests include Intelligent Transportation Systems (ITS), Architecture Design of CAVH System, Vehicle-infrastructure Cooperative Planning and Decision Method, Traffic Data Mining based on Cellular Data, and Traffic Operations and Management.
\end{IEEEbiography}

\vspace{11pt}

\vfill

\end{document}